\documentclass[useAMS,usenatbib,usegraphicx]{mn2e}
\usepackage{pdflscape}
\usepackage{multirow}
\usepackage{natbib}
\usepackage{aas_macros}
\usepackage{amsmath}

\newcommand{\gasp}{\texttt{GASP2D}}
\newcommand{\Sersic}{S\'ersic }
\newcommand{\galaxyz}{\texttt{galaXYZ}}

\title[Deconstructing double-barred galaxies: bulges]{Deconstructing double-barred galaxies in 2D and 3D. I. 
Classical nature of the dominant bulges} 

\author[A. de Lorenzo-C\'aceres et al.]{A. de Lorenzo-C\'aceres$^{1,2,3}$\thanks{E-mail:
adrianadelorenzocaceres@gmail.com}, J. M\'endez-Abreu$^{1,2,3}$, B. Thorne$^{1,4,5}$, and L. Costantin$^{6,7}$\\
$^{1}$School of Physics and Astronomy, University of St Andrews, North Haugh, KY16 9SS, Scotland, UK (SUPA)\\
$^{2}$Instituto de Astrof\'isica de Canarias (IAC), E-38205 La Laguna, Tenerife, Spain\\
$^{3}$Universidad de La Laguna (ULL), Departamento de Astrof\'isica, E-38206 La Laguna, Tenerife, Spain\\
$^{4}$University of Oxford, Denys Wilkinson Building, Keble Road, Oxford OX1 3RH, UK\\
$^{5}$Department of Astrophysical Sciences, Peyton Hall, Princeton University, Princeton, NJ, USA 08544\\
$^{6}$INAF Osservatorio Astronomico di Brera, via Brera, 28, 20159 Milano, Italy\\
$^{7}$Dipartimento di Fisica e Astronomia G. Galilei, Universita di Padova, vicolo dell Osservatorio 3, 35122 Padova, Italy
}

\begin{document}
\date{Accepted ***. Received ***; in original form ***}

\pagerange{\pageref{firstpage}--\pageref{lastpage}} \pubyear{2018}

\maketitle

\label{firstpage}

\begin{abstract}
We present here a thorough photometric analysis of double-barred galaxies, consisting of
i) two-dimensional photometric decompositions including a bulge, inner bar, outer bar, and 
(truncated) disc; and ii) three-dimensional statistical deprojections 
to derive the intrinsic shape of bulges, inner bars, and outer bars. 
This is the first time the combination of both techniques is applied to a sample
of double-barred galaxies.
It represents a step forward with respect to previous works,
which are based on properties of the integrated light through ellipse fitting and unsharp masking. 
In this first paper of a series of two, we analyse the nature of the dominant
bulges within double-barred systems 
by using several photometric diagnostics, namely
\Sersic index, Kormendy relation, colours, and the better suited intrinsic flattening. 
Our results indicate that almost all bulges in our sample are classical, whereas
only 2 out of the 17  galaxies under study appear as potential candidates to host 
secularly-formed disc-like bulges.
Such result poses the possibility that having a central hot structure may be a requirement for 
inner bar formation.
\end{abstract}

\section{Introduction}\label{sec:intro}
Double-barred galaxies are disc-like galaxies hosting two stellar structures
in the shape of a bar: an outer bar alike to the single
bars observed in $\sim$60-70\% of all disc galaxies \citep{Aguerrietal2009, Erwin2018}, 
and the so-called inner bar that refers to
an additional, smaller barred structure that has been found in at least $\sim$30\% of all
barred galaxies \citep{Laineetal2002, Erwin2004}.

Observations of galaxies with two nested stellar bars date back to the second half of the
twentieth century \citep[see][for a first identification of 
NGC\,1291 as a double-barred galaxy]{deVaucouleurs75}.
After serendipitous detections,
such as those of \citet{SandageandBrucato79} and \citet{Kormendy79}, 
\citet{Erwin2004} was the first to make a thorough compilation
of photometric data from various facilities in a catalog of 50 double-barred galaxies.
More individuals have been later detected by large photometric surveys
such as the Spitzer Survey of Stellar Structures in Galaxies \citep[S$^4$G; ][]{Shethetal2010}.
Moreover, inner bars have been observed up to a redshift 
$\sim$0.15 \citep{Liskeretal2006b}, thus supporting an scenario in which they are
not occasional but common stellar structures. This is a surprising result
as theoretically it is not possible to have two sets of closed barred orbits coexisting
within a rotating disc; the problem of the orbital support of double-barred systems
requires the complex definition of \emph{loops} introduced by \citet{MaciejewskiandSparke97}
and \citet{MaciejewskiandSparke2000}.
Simulating the formation and evolution
of double-barred galaxies has indeed been proved a hard task \citep[e.g.][]{Duetal2015,Wozniak2015}.

After almost 40 years of studies, many important questions remain open about double-barred galaxies. 
This work aims at providing photometric constraints to some of them through the most complete photometric study 
of double-barred galaxies ever performed. Our analysis
combines two-dimensional (2D) photometric decompositions of double-barred galaxies
in their multiple structural components (bulges, bars, and discs), and 
a three-dimensional (3D) statistical deprojection of their bulges and (inner and outer) bars.
We remark this is the first time either 2D decompositions or 3D deprojections of a sample of double-barred
galaxies are presented. 

In a series of two papers, we explore the following unanswered questions about double-barred galaxies:
i) whether there exists a major incidence of disc-like bulges within
double-barred galaxies, where secular evolution is assumed to take place in a very
efficient way; ii) whether inner bars form secularly out of disc-like bulges
already present in barred galaxies; iii) whether inner bars are transient or long-lived structures;
and iv) whether all barred galaxies will develop an inner bar at some stage of their lives.
This first paper is devoted to the presentation of the analysis and 
the photometric properties of bulges within double-barred systems (questions i and ii), while
the  properties of inner and outer bars 
(questions iii and iv) are 
studied in de Lorenzo-C\'aceres et al. (2018c, in preparation; hereafter Paper II).

\subsection{Frequency of disc-like bulges in double-barred galaxies}
Galactic bulges are properly defined as the central excess of light found 
in the surface brightness distribution of a disc galaxy, i.e., they imply 
a central concentration of stars. According to their photometric 
and kinematic properties, these structures are classified into two main groups:
classical and disc-like bulges. Both types
are believed to be linked to the different evolutionary processes forming them.
Classical bulges are considered the result of fast actions, such 
as monolithic collapse or mergers 
\citep[][]{Eggenetal1962,Aguerrietal2001, Bournaudetal2007,Hopkinsetal2010}.
These are pressure-supported systems rather analogous to elliptical galaxies.
On the other hand, disc-like bulges show more ordered motions dominated by
rotation. Sometimes referred to as pseudobulges, they are thought to be formed
through secular evolution mainly driven by bars
\citep[][although see \citealt{ElicheMoraletal2006} for a numerical evidence
of minor mergers causing similar bulge growth]{KormendyandKennicutt2004}.
Indeed, bars promote an angular momentum exchange between 
dark and baryonic matter or even between baryons and baryons. As a consequence,
stars are dragged
towards the outer galactic regions or gas is brought into the centre, where it concentrates
and may eventually form new stars and stellar structures such as bulges
\citep{Combesetal90, FriedliandBenz95, MunozTunonetal2004, Booneetal2007}.

\citet{Shlosmanetal89,Shlosmanetal90} show that a double-barred system
may theoretically be more efficient than a single bar 
in the transportation of gas towards the central
galactic regions, being even able to reach the sphere of influence of the central black hole. 
This is the reason why double bars have been proposed as
the gas channel for triggering
active galactic nuclei, although no conclusive observational evidence has been found
so far \citep{Marquezetal2000}. It is therefore sensible to state that
a major incidence of secularly-formed structures should be found in double-barred galaxies,
where gas inflow is very efficient. Disc-like bulges are commonly considered
the clearest consequence of secular evolution, and
inner bars themselves are  
considered as evidence for the presence of secularly-promoted bulges by many authors
\citep{FisherandDrory2016}.
Whether disc-like bulges are frequent in double-barred galaxies has not been 
purposedly studied in the literature so far, with the exception of 
\citet{deLorenzoCaceresetal2012}, who present
evidence of the existence of a disc-like bulge in the centre of the double-barred
galaxy NGC\,357. 

During the last decades, great effort has been put into the search of photometric diagnostics
that allow to discern the nature of bulges (i.e. classical vs. disc-like). The \Sersic index, $n$, appeared at first 
to be the best candidate. For example, \citet{FisherandDrory2008} found that classical 
bulges tend to have $n>$2, whereas disc-like bulges show $n<$2. 
The \Sersic index is a mathematical parameter describing the shape of the surface-brightness
profile of a bulge, once isolated from the rest of galaxy components through a photometric decomposition. Another diagnostic involves the 
projection of the fundamental plane onto the integrated surface 
brightness of the bulge within one effective radius versus the bulge effective
radius, also known as the \citet{Kormendy1977} relation. This 
has been argued to be a good discriminator between
classical and disc-like bulges, as it shows a correlation that creates a top sequence
mostly populated by classical bulges, while the existence of bulges with 
lower effective surface brightness for a given radius (i.e., outside the classical relation)
is explained by their disc-like nature
\citep[e.g.][]{FisherandDrory2008, Gadotti2009, FisherandDrory2016, Neumannetal2017}. 

More recently, \citet{Costantinetal2018b} tested the photometric diagnostics commonly used
to discern between classical and disc-like bulges,
and compared them with some kinematic diagnostics. Their results indicate that projected quantities
such as the \Sersic
index alone do not provide conclusive results and they propose the use of the intrinsic 3D shape
of bulges as a better photometric discriminator, based on the intrinsic flattening of the bulge component
with respect to the other galaxy structures \citep[see also][]{MendezAbreuetal2018}. 
Indeed, classical bulges are expected to be 
pretty spherical structures. On the other hand, 
rotating disc-like bulges should appear flatter, also depending on the
disc thickening.

\subsection{2D photometric decompositions}

All published photometric studies about double-barred galaxies so far rely 
on analyses of the integrated galaxy light \citep[e.g.][]{Erwin2004}.
Inner bars are usually detected through isophotal analysis as bumps in the ellipticity and position
angle profiles \citep[][]{FriedliandMartinet93, Laineetal2002}, 
and/or through a careful visual inspection of unsharp-masked images
\citep[][]{ErwinandSparke2003}. 
Better suited photometric decompositions, which model the galaxy light as a combination
of individual structures instead of studying the integrated properties,
have not been systematically applied for double-barred cases. The only notable exceptions are the
recent works by \citet{MendezAbreuetal2017}, who performed 2D photometric decompositions of all galaxies of the
Calar Alto Legacy Integral-Field Area survey \citep[CALIFA; ][]{Sanchezetal2012}, including
two double-barred cases, and 
de Lorenzo-C\'aceres et al. (2018a, submitted; see also 
\citealt{MendezAbreuetal2019}), 
who decomposed the S$^4$G images 
of the two double-barred galaxies present in the spectroscopic TIMER project
\citep{Gadottietal2018}.

The rapid progress in the field of photometric decompositions has resulted in 
several sophisticated codes that dissect the galaxy into various components
by making 2D models of each structure \citep[e.g.][]{Pengetal2002, deSouzaetal2004, MendezAbreuetal2008, Erwinetal2015b}.
However, most of the published works include up to two components, namely
disc and bulge, while other structures such as stellar bars are not taken 
into account \citep[e.g.,][]{Simardetal2011}. Bulge and disc certainly account for most of the
galaxy light but not including bars has an effect on the measured parameters,
as shown by \citet{Laurikainenetal2006}, \citet{Gadotti2008}, \citet{FisherandDrory2008}, 
and \citet{MendezAbreuetal2017}, among others. Dismissing the bar leads to
large uncertainties and overestimation of the bulge parameters ($n$ and effective radius)
and bulge contribution to total light. 

Here we present the first consistent 2D multi-component photometric decomposition
of a homogenous set of images of 17 double-barred galaxies,
fitting a bulge, a (truncated) disc, and the two bars with the code \gasp\
\citep{MendezAbreuetal2008}. The analysis of the bulge properties is therefore
performed with no contamination of the inner and outer bars.

\subsection{3D deprojection: intrinsic shapes}
While photometric decompositions provide the individual parameters for the structural
components shaping galaxies, such analysis is based on projected images of the galaxies on the sky.
Therefore, and even after deprojecting bar lengths and ellipticities, only the two dimensions
in the galaxy plane
are explored. Recent development in mathematical and statistical techniques
has shown that it is possible to retrieve information about the missing third direction and
actually derive the intrinsic 3D shape of galaxy structures (bulges and bars), as
reviewed in \citet{MendezAbreu2016}. Studies so far have focused on the case of non-barred and single-barred
galaxies. In the set of papers by \citet{Costantinetal2018a} and \citet{MendezAbreuetal2018b},
the intrinsic shapes of bulges and bars in the CALIFA sample are analysed.
Among other conclusions, the authors find that the intrinsic flattening of bulges
with respect to bars holds important clues for understanding the formation path 
of these structures. 

In this project we derive the 3D shape of bulges and (inner and outer) 
bars with the code \galaxyz\ \citep{MendezAbreuetal2010b, Costantinetal2018a}.
We emphasise that this is the first time the intrinsic shape of inner bars within double-barred
systems are studied.
 \\

The paper is organised as follows: the parent sample and images are
described in Sect.\,\ref{sec:data}, while a brief outline of \gasp\
and the procedure followed in this work
to perform 2D multi-component photometric
decompositions of double-barred galaxies are provided in Sect.\,\ref{sec:GASP2D}.
Section\,\ref{sec:3D} describes \galaxyz\ and the procedure used to derive the 3D 
intrinsic shape of bars and bulges. In Sect.\,\ref{sec:influence} we 
quantify the possible bias 
introduced in the bulge parameters by dismissing the presence of an inner bar.
Section \ref{sec:bulgenature} shows an analysis of the photometric properties
of bulges in double-barred galaxies and the discussion about their nature is provided
in Sect.\,\ref{sec:discussion}.
Conclusions are wrapped up in Sect.\,\ref{sec:conclusions}.
A flat cosmology with $\Omega_{m}=$0.3, $\Omega_{\Lambda}=$0.7, and $H_0=$75\,km\,s$^{-1}$\,Mpc$^{-1}$ is 
assumed throughout the paper.
These are the same parameters adopted by \citet{Gadotti2009} and 
\citet{MendezAbreuetal2017}, whose works are used 
for comparison throughout this paper.

\section[]{The sample of double-barred galaxies}\label{sec:data}

The sample of double-barred  galaxies is extracted  from a catalog of
67 barred galaxies with inner structures presented in 
 \citet{Erwin2004}. Among  these,  50  galaxies  are
double-barred, as  classified by  two photometric  diagnostics, namely
ellipse fitting and unsharp masking.  Bar properties such as sizes and
position angles (both measured from ellipse fitting) are also provided
by  \citet{Erwin2004}, who  uses  a compilation  of  results from  the
literature and  new measurements made  on a  variety of images  in the
optical and near infrared.

For the present analysis we  first selected all double-barred galaxies
from the catalog of \citet{Erwin2004} with available Sloan Digital Sky
Survey  \citep[SDSS;  ][]{Yorketal2000}   imaging,  thus  obtaining  a
preliminary sample of 23 objects.  SDSS provides a homogeneous set of
$g'$-, $r'$-, and $i'$-band images with medium spatial resolution, suitable
for inner bar  detection at the redshift of  our galaxies ($z<0.015$),
and  a field-of-view  large enough  to  reach the  outermost
  regions of the galaxy, as  required  for  a proper  modelling  of the  disc
component. In particular, we use the images from the SDSS Data Release
9  \citep{Ahnetal12}. All  images  are  already soft-bias  subtracted,
flat-field corrected,  sky subtracted,  and flux calibrated  using the
standard SDSS pipelines.  Some further treatment  of the data
is necessary to  both re-calibrate the images  from nanomaggies to
counts (required for  the fitting procedure with \gasp)  and to refine
the  sky  subtraction.   Details  on this  process  are  described  in
\citet{Pagottoetal2017} and  \citet{Costantinetal2018b}. The
point  spread  function (PSF)  is  measured  on  each image  using  a
circular Moffat  function \citep{Moffat1969}.  The mean  values of the
Full  Width  at  Half  Maximum   (FWHM)  for  the  $g'$-,  $r'$-,  and
$i'$-images are 1.32, 1.15, and 1.17\,arcsec, respectively.

Notwithstanding the careful inspection  made by \citet{Erwin2004}, our
2D  multi-component  photometric decompositions  reveal
that  some  of  these  galaxies   had  been  either  misclassified  as
double barred or the SDSS spatial  resolution is not enough to readily
distinguish the inner bar. In the former case, we found this is mainly
due to the presence of other  components (e.g., stellar inner rings or
complex  dust structures).   
Six galaxies were  finally removed
from our  preliminary sample, namely Mrk\,573,  UGC\,524, NGC\,1068,
NGC\,4303, NGC\,4321,  and NGC\,4736.   Our definitive sample  is therefore
composed of  17 double-barred galaxies.  Table  \ref{tab:sample} shows
the galaxy sample together with some relevant parameters.

All  our double-barred  galaxies are nearby,  with $z<0.015$  as
indicated   in  Table   \ref{tab:sample}.   \citet{FisherandDrory2016}
establish that, given the SDSS  spatial resolution and a typical bulge
effective  radius  of  2\,kpc, photometric  decompositions  aiming  at
accurately deriving bulge properties with SDSS data should restrict to
galaxies  up  to  $z$=0.03.   \citet{Costantinetal2017}  explore  the
possible errors on the bulge  structural parameters when their angular
sizes are close to the size of the image PSF. They find that even for
bulges with  effective radius 1.2$\times\sigma$ of the  PSF ($\sigma
\sim \rm FWHM/2.35$), the bulge parameters  can be recovered within a 10\%
error.  We conclude  that  our photometric  decomposition analysis  is
therefore  not hampered  by resolution  effects
affecting small bulges or inner bars.

\begin{table}
 \centering
  \caption{DOUBLE-BARRED GALAXIES SAMPLE.}
\resizebox{8cm}{!}{  \label{tab:sample}
  \begin{tabular}{lccc}
  \hline
Name  & Morphological type & Distance (Mpc) & $z$\\
 (1)  & (2) & (3) & (4)\\
\hline
\hline
NGC\,357  & SB(r)0/a        & 31.6 & 0.008\\
NGC\,718  & SAB(s)a         & 22.6 & 0.006\\
NGC\,2642 & SB(r)bc         & 56.8 & 0.014\\
NGC\,2681 & (R')SAB(rs)0/a  & 17.2 & 0.002\\
NGC\,2859 & (R)SB(r)0$^+$   & 24.3 & 0.006\\
NGC\,2950 & (R)SB(r)0$^0$   & 14.9 & 0.004\\
NGC\,2962 & (R)SAB(rs)0$^+$ & 30.0 & 0.007\\
NGC\,3368 & SAB(rs)ab       & 10.5 & 0.003\\
NGC\,3941 & SB(s)0$^0$      & 12.2 & 0.003\\
NGC\,3945 & (R)SB(rs)0$^+$  & 19.3 & 0.004 \\
NGC\,4314 & SB(rs)a         & 12.0 & 0.003\\ 
NGC\,4340 & SB(r)0$^+$      & 15.3 & 0.003\\
NGC\,4503 & SB0$^-$:        & 15.3 & 0.004\\
NGC\,4725 & SAB(r)ab        & 12.4 & 0.004\\
NGC\,5850 & SB(r)b          & 35.2 & 0.009\\
NGC\,7280 & SAB(r)0$^+$     & 24.3 & 0.006\\
NGC\,7716 & SAB(r)b:        & 34.1 & 0.009\\
\hline
\end{tabular}
}
\begin{minipage}{7cm}
Notes. (1) Galaxy name; (2) and (3) morphological types and luminosity distances as extracted from the catalog of \citet{Erwin2004}.
Distances are corrected from Virgocentric motion; 
(4) redshifts from NED.
\end{minipage}\end{table}

\subsection{Control samples of single-barred galaxies}\label{sub:control}
Two control samples of single-barred galaxies are used throughout this paper: those
from \citet{MendezAbreuetal2017} and \citet{Gadotti2009}.

\citet{MendezAbreuetal2017} present the 2D photometric decompositions
of the SDSS $g'-$, $r'-$, and $i'-$images for 404 galaxies from the 
CALIFA survey \citep[][]{Sanchezetal2012}.
They make use of the \gasp\ code.
The methodology used in \citet{MendezAbreuetal2017} is very similar to the one followed in this work and 
described in Sect.\,\ref{sec:GASP2D}. 
For this reason, the photometric parameters 
for their 162 single-barred galaxies are used as a control sample throughout this paper.
Note that the CALIFA sample expands a wider redshift range than our double-barred sample, covering from
$z$=0.005 up to $z$=0.03. In addition, the analysis of \citet{MendezAbreuetal2017}  includes two
double-bar hosts as well: NGC\,7716 and NGC\,23. While NGC\,7716 is included in the sample presented here, 
NGC\,23 does not belong to the parent sample from \citet{Erwin2004} and it was therefore not 
initially considered for this work. Results from \citet{MendezAbreuetal2017} for these two galaxies
are shown together with our measurements whenever it is possible, 
in order to demonstrate the good consistency between both works and particularly
for the one galaxy in common. 

\citet{Gadotti2009} also performs 2D multi-component photometric decomposition of SDSS
images in all the bands. 
His analysis includes nearly 1000 galaxies, among which he finds 287 bar+bulge hosts.
The sample is more distant (0.02$<z<$0.07) than the sample presented here,
and no search for possible double bars is performed; 
some level of contamination
is therefore expected. Note that the code used for the analysis 
of \citet{Gadotti2009} is {\texttt{BUDDA}} \citep{deSouzaetal2004},
which describes bars with \Sersic profiles instead of the Ferrers 
profiles used in this work (see Sect.\,\ref{sec:GASP2D}). By using the works of 
both \citet{MendezAbreuetal2017} and \citet{Gadotti2009} as control single-barred
samples for our analysis, we further assess possible biases due to
the different approaches considered by the fitting codes.

\section[]{Two-dimensional multi-component photometric decompositions with \gasp}\label{sec:GASP2D}

The  2D multi-component  photometric decompositions  are
performed    with   the    code   \gasp\    \citep{MendezAbreuetal2008,
  MendezAbreuetal2014,  MendezAbreuetal2017}.  \gasp\  fits the  galaxy
2D surface-brightness distribution  with a combination of
structural components, parameterised  by known mathematical functions.
A  Levenberg-Marquardt algorithm based on a $\chi^2$ minimisation is used  to find  the most  suitable
set of parameters describing the galaxy light.

\gasp\  has already  been  tested in  a  number of  works  to find  the
structural  composition of  a variety  of objects,  such as  AGN hosts
\citep{Benitezetal2013},   galaxies  with   decoupled  polar bulges
\citep{corsinietal2012},         and         isolated         galaxies
\citep{morellietal2016}.   For  the  present  study,  the  ability  of
simultaneously fitting  two bar structures  was added to the  code, as
already  introduced  in  \citet{MendezAbreuetal2017}.   The  available
components for the fitting are therefore  a bulge, up to two bars, and
a disc that might show none, positive, or negative bending
\citep[e.g.][]{Erwinetal2005, PohlenandTrujillo2006}.
 
For the  sake of completeness,  in the  following we will  present the
analytical functions  describing each component.  More  details on the
fitting procedure can be found in \citet{MendezAbreuetal2017}.


The  surface  brightness  distribution   of  the  bulge  component  is
parameterised with a \Sersic  profile \citep{Sersic68} of the form

\begin{equation} 
I_{\rm b}(r_{\rm b})=I_{\rm e}10^{-b_n\left[\left(\frac{r_{\rm b}}{R_{\rm e}} 
\right)^{\frac{1}{n}}-1\right]}, 
\label{eqn:bulge_surfbright} 
\end{equation} 
%
where $r_{\rm  b}$ is the radius  measured in the reference  system of
the bulge.   $R_{\rm e}$, $I_{\rm e}$,  and $n$ are the  effective (or
half-light) radius,  the surface  brightness at  $R_{\rm e}$,  and the
\Sersic index describing the  curvature of the profile, respectively,
and $b_n \simeq 0.868\,n-0.142$ \citep{Caonetal93}.

The surface  brightness distribution of  a galaxy disc is  allowed to
take  three different  shapes, namely:  (i) Type  I profile,  a single
exponential profile,  (ii) Type II  profile, a double  exponential law
with a down-bending beyond the  so-called break radius, and (iii) Type
III profile, a double exponential law  with an up-bending in the outer
parts of the  disc.  To account for these possibilities  we adopt the
following parameterisation:

\begin{equation} 
I_{\rm d}(r_{\rm d})=I_{\rm 0}\, \left[e^{\frac{-r_{\rm d}}{h}}\, \theta \, + \, e^{\frac{-r_{\rm break}\,(h_{\rm out}-h)}{h_{\rm out}\,h}}\, e^{\frac{-r_{\rm d}}{h_{\rm out}}}\,(1-\theta)\right] ,
\label{eqn:disk_trunc} 
\end{equation} 
%
where
\[
\theta =
\begin{cases}
   0  & \qquad {\rm if} \qquad r_{\rm d} > r_{\rm break} \nonumber \\
   1  & \qquad {\rm if} \qquad r_{\rm d} < r_{\rm break},
\end{cases}
\]
%
and $r_{\rm d}$ is the radius  measured in the reference system of the
disc.  $I_0$, $h$, $h_{\rm out}$,  and $r_{\rm break}$ are the central
surface brightness, inner scale-length,  outer scale-length, and break
radius of the disc, respectively.

The projected surface density of a 3D Ferrers ellipsoid
(\citealt{Ferrers77}, see  also \citealt{Aguerrietal2009}) is  used to
describe the surface-brightness distribution of both the inner and outer bar components:

\begin{equation}
I_{\rm bar}(r_{\rm bar})=I_{\rm 0,bar}\left[1-\left(\frac{r_{\rm bar}}{a_{\rm bar}}\right)^2\right]^{n_{\rm bar}+0.5}; 
\qquad {\rm for}\, r_{\rm bar} \le a_{\rm bar},
\end{equation}
%
where $r_{\rm  bar }$ is  the radius in  the reference system  of each
bar.  The  inner   and  outer  bars  are  allowed   to  have  different
ellipticities and  position angles.   $I_{\rm 0,bar}$,  $a_{\rm bar}$,
and $n_{\rm  bar}$ represent  the central surface  brightness, length,
and shape parameter of the bar,  respectively. It is worth noting that
$a_{\rm bar }$ is not an effective radius, but  the radius where
the bar intensity drops to zero.

The  bar surface-brightness  distribution  is assumed  to be  axially
symmetric     with     respect     to    a     generalised     ellipse
\citep{athanassoula90}. Therefore, the radial coordinate is defined as
%
\begin{equation} 
r= \left( |x|^c + 
\left| \frac{y}{(1-\epsilon_{\rm bar})}\right|^c \right)^{1/c}, 
\label{eq:general_ellipse}
\end{equation} 
%
where $\epsilon_{\rm bar}$ is the bar ellipticity and $c$ controls the
shape  of the  isophotes. A  bar  with pure  elliptical isophotes  has
$c$=2. It is  $c>$2 if the isophotes  are boxy, and $c<$2  if they are
discy.  The  parameters $\epsilon_{\rm bar}$  and $c$, as well  as the
position angle, are assumed to be constant as a function of radius.

\subsection[]{Selection of the $n$ and $c$ bar parameters}\label{sub:nandc}

\begin{figure*}
 \vspace{2pt}
 \includegraphics[angle=0., width=0.8\textwidth]{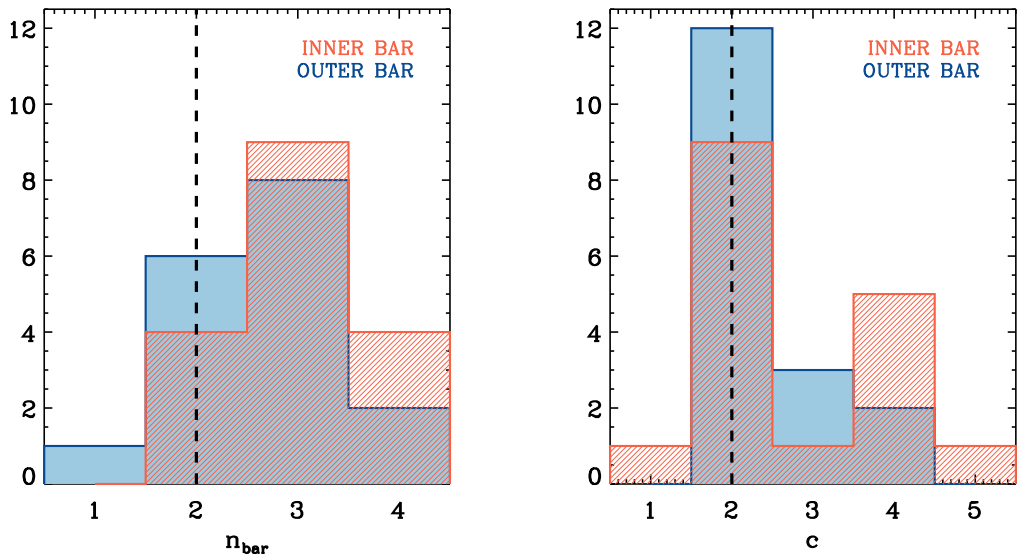}
 \caption{Distribution of the $n_{\rm bar}$ and $c$ bar parameters as calculated by the procedure described 
in Sect.\,\ref{sub:nandc}. Blue and orange histograms correspond to outer and inner bars, respectively.
The vertical dashed lines mark the position of the default values commonly adopted in the literature. While
most bars, especially outer bars, are well reproduced with the standard $c$=2, 
$n_{\rm bar}$ shows a wider distribution.}
 \label{fig:nandc}
\end{figure*}

%
%
Recovering  all  possible bar  parameters 
previously described as free variables in  the
fitting process is rather  difficult even for  single-barred galaxies,
due to the high number of degeneracies among the parameters.  The most commonly used
procedure in photometric decompositions  involving Ferrers profiles is
to keep fixed the two shape parameters to their  default $n_{\rm bar}$=2
and $c$=2 values \citep{Laurikainenetal2005, MendezAbreuetal2017}.

Since  we are  particularly interested  in studying  the structure  of
inner and outer  bars with great accuracy, we  have investigated which
values of  $n_{\rm bar}$ and $c$  provide the best  fits.  For this  purpose, we
first perform the $r'$-band double-barred  fit in the usual way, i.e.,
with fixed values of $n_{\rm bar}$=2 and $c$=2. Variations of  the profile with $n_{\rm bar}$ are explored
first: the results from the usual fits are introduced as fixed initial
conditions for all the inner  and outer bar parameters ($I_{\rm 0,bar}$, $a_{\rm bar}$, 
$b/a$,  and $PA$) except for  the outer bar length $a_{\rm OB}$,  which is
allowed to vary. We remark that the correlation between bar length and
$n_{\rm bar}$ makes it mandatory to keep the  bar length as a free variable when
studying  variations of  $n_{\rm bar}$.   \gasp\  is then  run  again with  fixed
integer  values for $n_{\rm OB}\in[1,4]$. The trends  $n_{\rm OB}$ vs
$\chi^2$ are inspected so the minimum providing the best $n_{\rm OB}$
parameter for the outer bar is found. The process is then repeated for
the  inner bar  case, fixing  both  $n_{\rm OB}$ and  $a_{\rm OB}$
parameters to the newly recovered values.

A similar  procedure is  carried out  to derive  the best  $c$ values,
varying the integer  values of $c$ within the  range $c\in[1,5]$.  
For this case, bar length as  well as the rest
of bar  parameters (including the  updated $n_{\rm bar}$ values just  obtained in
the previous step)  can be kept fixed and therefore  this procedure is
just a  $\chi^2$ computation rather  than a fitting. Again,  the outer
bar $c$ is explored first.

As the inner bar  is small and located at the  galaxy centre, a slight
modification of  its global shape does  not affect the large  scale of
the outer  bar fitting.  However,  the extrapolation of the  outer bar
profile towards the central regions accounts  for a certain amount of light
that     may     significantly     affect     the     inner     bar
contribution. Determining the  outer bar shape in  first place, before
exploring $n_{\rm bar}$  and $c$ variations  for the  inner bar, is  therefore a
sensible procedure.

The   final    parameters   for   the    2D  photometric
decompositions of  double-barred galaxies in $r'$-band,  including $n_{\rm bar}$
and $c$  values for the Ferrers  bar profiles, are included  in Tables\,\ref{tab:paramsr1}   
and  \ref{tab:paramsr3}   and   shown  in   Fig.\,\ref{fig:nandc}.  
We find  that two thirds of the outer  bars are well
reproduced  with  the  default  value  $c$=2,  while  this  percentage
slightly decreases  down to 53\% for  the inner bar case.   The remaining inner
and outer  bars tend to be  boxier than the nominal  elliptical case.
$n_{\rm bar}$ values differ  even more from the  standard case, as only  6 and 4
out of 17 outer and inner bars, respectively, show $n_{\rm bar}$=2. Except for one
outer  bar, bars  in  general show  steeper profiles  with
$n_{\rm bar}>$2, being $n_{\rm bar}$=3 the preferred value for most cases.

\subsection[]{Double- and single-bar fits}\label{sub:singleanddouble}
\begin{figure*}
 \vspace{2pt}
\centering
 \includegraphics[angle=270., width=1.\textwidth]{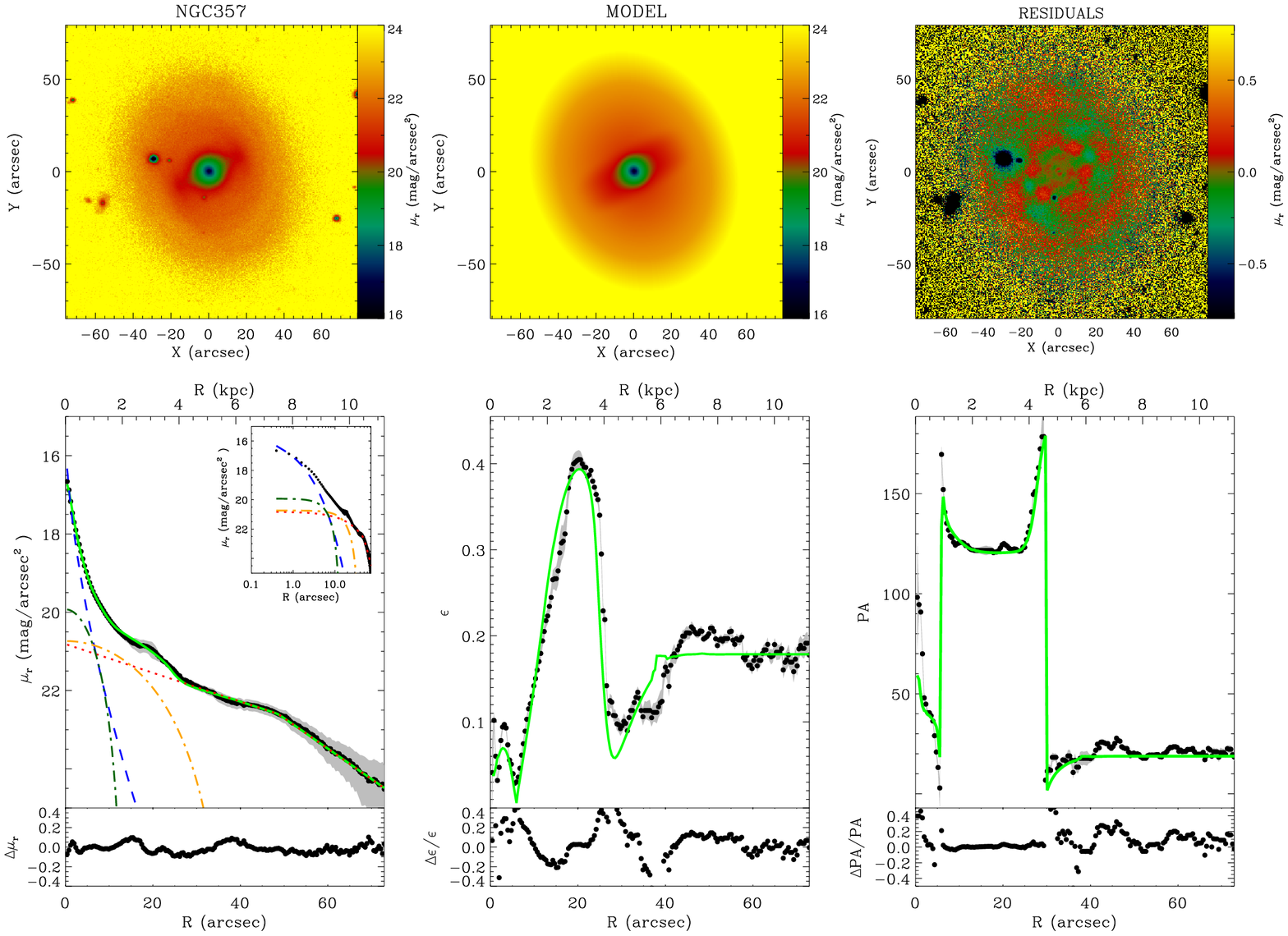}
 \caption{2D multi-component photometric decomposition of the SDSS-DR9 $r'$-band image of the 
double-barred galaxy NGC\,357 performed with \gasp. The top panels show the original image (left),
the 2D best-fitting model (middle), and the residuals (right). The bottom left panel
shows the original surface-brightness radial profile (black points) as derived with an isophotal fitting.
Blue dashed, green dash-dotted, yellow dash-dotted, and red dotted lines show the bulge, inner bar, outer
bar, and disc components, respectively. The disc is truncated showing a down-bending. Residuals are
included in the lower subpanel while the inset zooms into the very central region. Bottom middle and bottom right
panels show the ellipticity and position angle profiles from the isophotal fitting of the original image
(black dots) and the same measurements for the 2D model (green lines).}
 \label{fig:fitexample}
\end{figure*}

All the  sample galaxies  were found  to host  a bulge  component, two
bars, and  a disc.   The $r'$-band  images are  used as  benchmarks to
perform the  first fit.  Figure\,\ref{fig:fitexample}  shows an example
of the final $r'$-band fit for  the galaxy NGC\,357, which is composed
of a Type-II disc, a bulge,  and the inner and outer bars.  

Following the  prescription given in  \citet{MendezAbreuetal2017}, the
final parameters  for the  $r'$-band are used  as initial  guesses for
fitting the $g'$- and $i'$-band images,  fixing the $n_{\rm bar}$ and $c$ values
to   the   best   estimates   obtained   as   described   in   Sect.\,\ref{sub:nandc}.    
Although  very   similar,  slight   band-dependent
differences in  the measured  parameters are expected.   Best-fitting
parameters for all the structures in each galaxy and band are shown in
Tables\,\ref{tab:paramsr1} to \ref{tab:paramsi3}.
%

The results from the double-barred fits in the $r'$-band are also used
as initial guesses for single-barred  $r'$-band fits. The outcomes are
found  in   Tables\,\ref{tab:paramsone1} and
\ref{tab:paramsone3}.    These  single-barred   fits   are  done   for
comparison purposes as discussed in Sect.\,\ref{sec:influence}.

We remark here that all parameters included in Tables\,\ref{tab:paramsr1} to
\ref{tab:paramsone3} correspond to
direct measurements from the images, i.e., lengths, radii, and scalelengths are provided
in arcsec and projected onto the plane of the sky. For the analysis presented
throughout the paper, bar parameters (lengths, ellipticities, and position
angles) have been deprojected following the recipes given by
\citet{Gadottietal2007} and angular sizes have been transformed into physical
scales. As explained in \citet{Zouetal2014},
bar parameters for galaxies with inclinations $i>$60$^{\circ}$ have not been deprojected
due to the high uncertinties introduced in the process.

\subsection{Error computation}\label{sub:errors}

The  errors on  the individual  parameters  involved in  the fit  were
computed using  a set of  tailor-made mock  galaxies in a  Monte Carlo
fashion.   The full  description of  the methodology  is presented  in
\citet{MendezAbreuetal2017} and we refer the  reader to this paper for
a complete description.  Here we provide  a brief summary for the sake
of clarity.

A sample  of 500  mock double-barred galaxies  was simulated  using a
combination of structural parameters  constrained within the limits of
our    real   galaxy    sample   in    the   $r'$-band    (see   
Tables\,\ref{tab:paramsr1} and \ref{tab:paramsr3}).   
Each model  was built  up using  the equations
provided in  Sect.\,\ref{sec:GASP2D} for  each distinct structure  on a
2D grid with the SDSS  pixel scale (0.396\,arcsec/px). The
total galaxy model  was then convolved with a circular  Moffat PSF with
the    typical    FWHM   of    our    SDSS    images   in    $r'$-band
(Sect.\,\ref{sec:GASP2D})  in order  to reproduce  the observed  spatial
resolution.  We also  adopted the  typical  values of  CCD gain  
(4.86\,e$^-$/ADU) and read-out  noise (5.76\,e$^-$), and  added the background
and  photon noise  from the  galaxy to  yield a  signal-to-noise ratio
similar to that of the observations.

Finally, the mock images were analysed  using \gasp\ in the same way as
real  images. The  difference  between the  input  and output  values
provides us with  a systematic (mean value)  and statistical (standard
deviation) error on the individual parameters. Both errors are added
in    quadrature   to    obtain    the   final    values   shown    in
Tables\,\ref{tab:paramsr1} to \ref{tab:paramsone3}.

\section[]{Three-dimensional shapes with \galaxyz}\label{sec:3D}
\begin{figure*}
 \vspace{2pt}
 \includegraphics[bb=54 335 558 720, angle=0., width=0.6\textwidth]{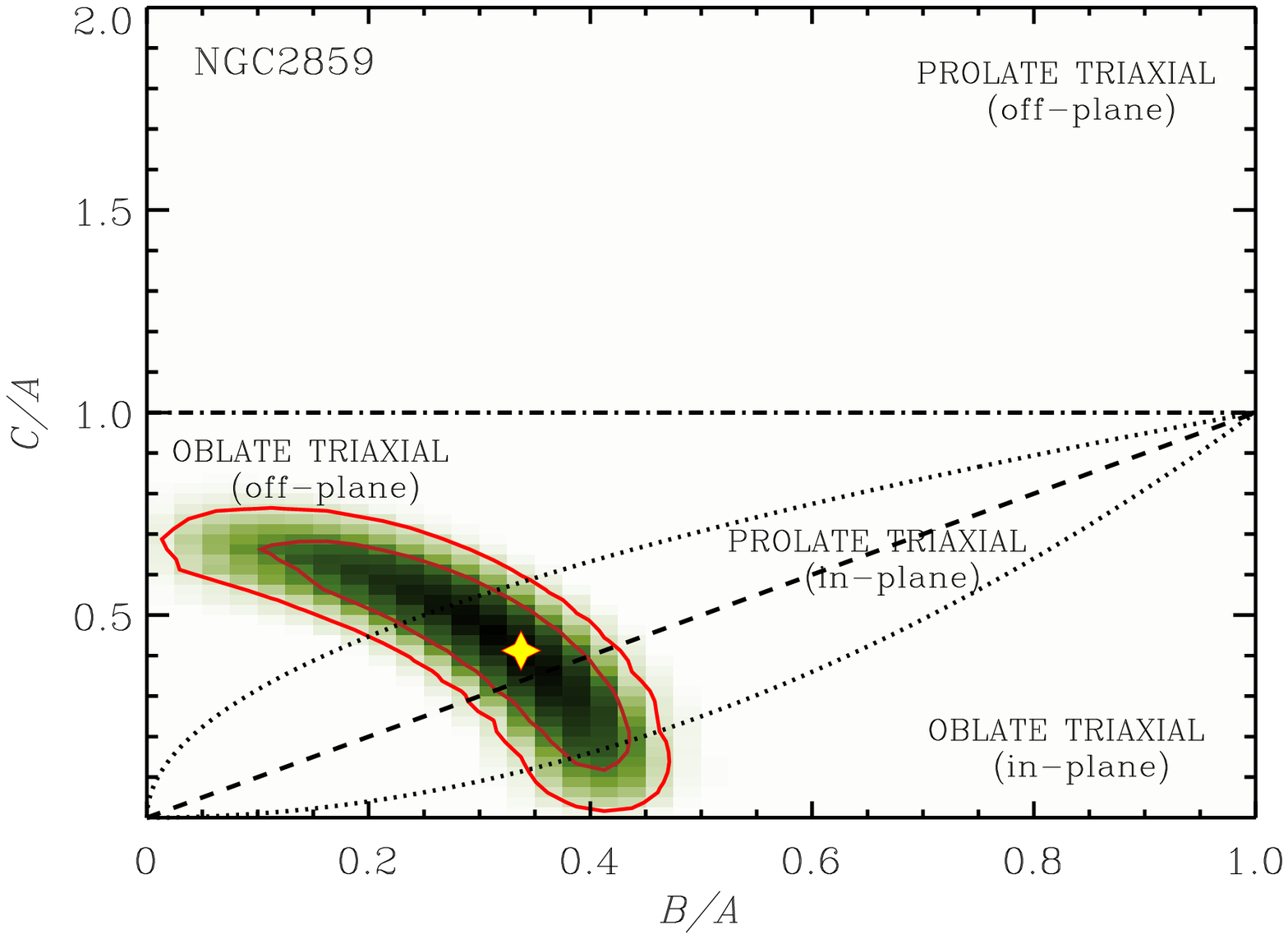}
 \caption{3D statistical derivation of the intrinsic shape for the inner bar of the double-barred
galaxy NGC\,2859. The joined probability distribution functions (PDF) of the in-plane and off-plane 
axis ratios, $B/A$ and $C/A$, are plotted in green colours. The value with the highest
probability is shown with a yellow diamond. The regions corresponding to different shapes are indicated
\citep[prolate, oblate, and triaxial; see][]{MendezAbreuetal2018b}. 
Such analysis has been performed for all bulges, inner bars, and outer bars
of the sample.}
 \label{fig:example3D}
\end{figure*}
\begin{figure*}
 \vspace{2pt}
 \includegraphics[angle=0., width=0.8\textwidth]{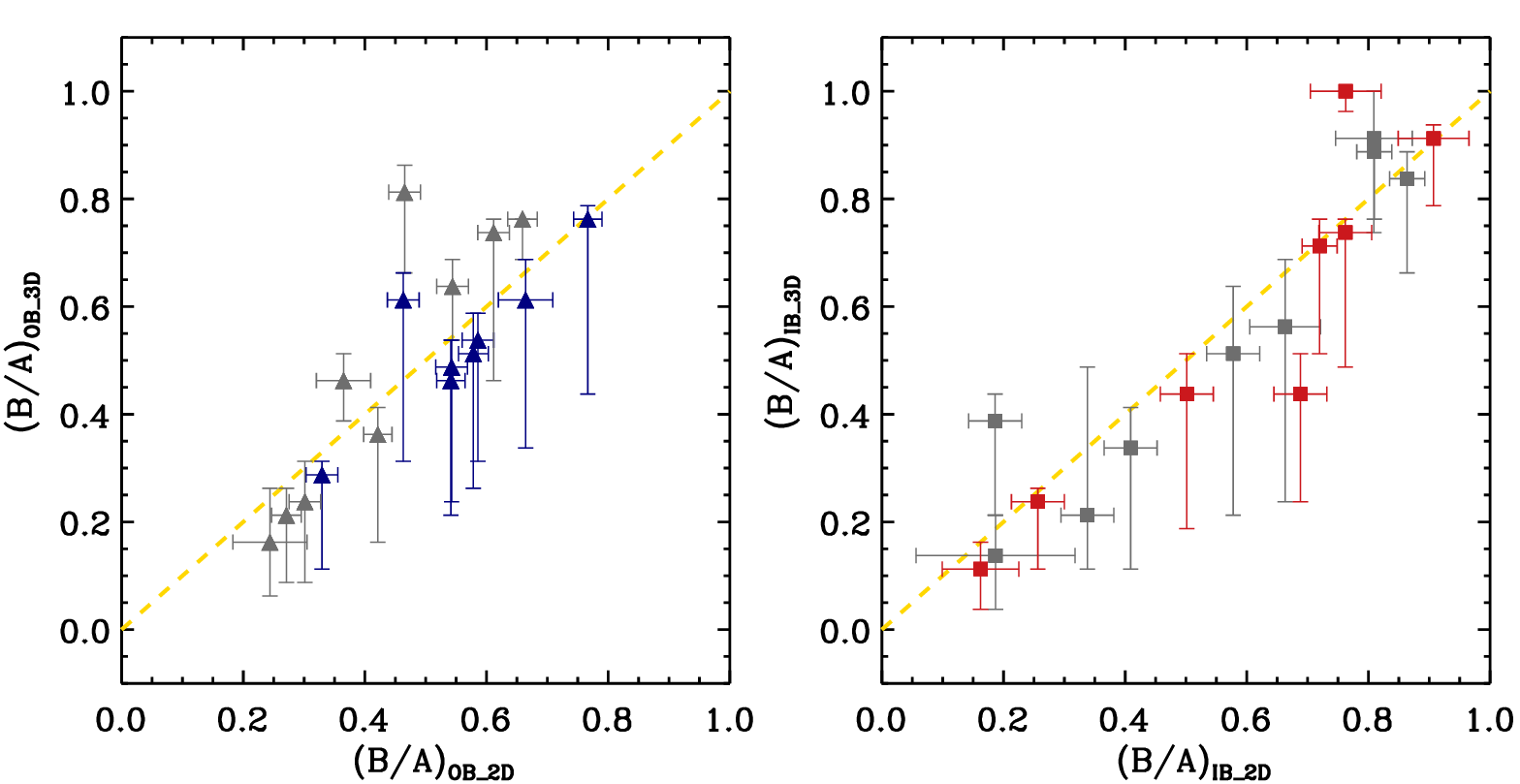}
 \caption{Comparison between the semi-major axis ratios in the galaxy plane 
for the outer (triangles; left) and inner (squares; right) bars as measured in 2D
with \gasp\ and then deprojected (horizontal axes), and measured in 3D with \galaxyz\ (which provides already deprojected
values; vertical axes). Coloured symbols correspond to galaxies with uncertainties in any semi-major axis 
($B/A$ or $C/A$ - not shown in this plot- for either the inner or
outer bar) less than 0.5, while the grey symbols represent
galaxies with larger uncertainties.
The good agreement between both techniques is indicated by a relationship close to the
yellow-dashed 1:1 line in each plot.}
 \label{fig:comparison2D3D}
\end{figure*}

We derive the intrinsic 3D
shape for the  bulges, outer bars, and inner bars  of our sample using
the \galaxyz\ code, which follows the procedure described in 
\citet{MendezAbreuetal2010b} and \citet{Costantinetal2018a}. This method has
been previously  applied to different  samples of galactic  bulges and
outer bars,  but it  is used here  for the first  time in  inner bars.
\citet{MendezAbreuetal2018b} and \citet{Costantinetal2018a} 
have  already  demonstrated  that  the  statistical
approach  to  derive the  intrinsic  3D  shape  is applicable  to  any
galactic   structure  if   the  initial   assumptions  are   fulfilled, namely:
i)  all structures under study  
can be  modelled by a triaxial ellipsoid in  the same equilibrium
plane as the disc;  ii) the galaxy disc is considered  to be an oblate
spheroid.   We  allow for  the  disc  to  have a  intrinsic  thickness
according to  a normal  distribution with  mean intrinsic  axial ratio
$\langle   q_{0,d}   \rangle$=0.267   and   standard   deviation
$\sigma_{q_{0, {\rm d}}}$=0.102 \citep{RodriguezandPadilla2013}; and iii) all
structures share  the same centre, which  is adopted as the  centre of
the  galaxy. 

All  previous conditions  are met  by the  three structures
studied here. A caveat  is that stellar bars  can develop
vertically-extended  components during  the so-called  buckling phase.
During this time,  the bar creates what is called  a box/peanut (B/P)
structure           in          its           central          regions
\citep{CombesandSanders81,Athanassoulaetal83,MartinezValpuestaetal2006}.      These
structures  do   not  comply with   our  first  hypothesis.    However,  in
\citet{MendezAbreuetal2018b}  it is demonstrated  that the  parameters obtained
from our photometric decomposition represent the
'thin' part of the bar. Possible B/P structures present in a
bar therefore have a small impact in the derived intrinsic shape.

The \galaxyz\ code  needs as input the measured values  of the projected
geometric parameters (ellipticity and position angle)
of the disc (representative of the galaxy inclination and position of the lines-of-nodes),
and the structures under study, i.e. bulges, outer bars, and inner bars in our case.
All these parameters,  and their
corresponding   errors,  are   provided by  the  2D   photometric
decomposition  carried out  in  Sect.\,\ref{sec:GASP2D}.  Then, \galaxyz\ randomly
generates 1000 geometric configurations by adopting for each parameter
a  Gaussian distribution  centred on  its measured  value and  with a
standard  deviation  equal to  its  uncertainty.   For each  geometric
configuration, the code evaluates equations 5 and 6 in \citet{Costantinetal2018a}. 
This is carried out in a Monte Carlo fashion with 5000 simulations of
the intrinsic semiaxis ratios $B/A$ and $C/A$.

The  result of  this  analysis is  a  joined probability  distribution
function (PDF) of  $B/A$ and $C/A$, i.e., the PDF  of the intrinsic 3D
shape of each structure. An example is shown in Fig.\,\ref{fig:example3D}
for the inner bar of NGC\,2859. Our approach is entirely
based on the projected geometric  properties of each structure and the
PDF is calculated  independently for each structure in  a galaxy.  The
most probable $B/A$ and $C/A$ values and their corresponding 1$\sigma$
uncertainties are  shown in Table\,\ref{tab:intrinsic}.   We notice here that
the width of the PDF in either  $B/A$ or $C/A$ does not only depend on
the  photometric  decomposition errors,  but  mostly  on the  lack  of
knowledge of  the Euler angle ($\phi$), i.e., the angle  describing the
position of the intrinsic major axis  of the structure with respect to
the line-of-nodes in the plane of the disc.  Therefore, there are some projected
configurations (combination  of the disc and  structure geometry) that
are less suited to derive the 3D shape and they provide large
uncertainties. Values with 1$\sigma$ errors  
in any  intrinsic semiaxis
ratio larger than 0.5 are not included in the final analysis.

We remark that \gasp\ provides the size of each structure 
as projected on the sky; those values are then deprojected following
the methodology described in \citet{Gadotti2009}, thus
sizes in the galaxy plane are finally estimated. On the other hand, \galaxyz\
performs a statistical recovery of the size of each structure in both
the galaxy plane and perpendicular to it.
Since \galaxyz\ uses the outcomes from \gasp\ as input values,
the results from both analyses are not fully independent. However, the PDF 
delivered by \galaxyz\ include statistical as well as methodology-intrinsic 
uncertainties, and we select as final value the one
single point with the maximun likelihood. A good agreement
between the in-plane sizes retrieved by \gasp\ and \galaxyz\ therefore
indicates the goodness of our combined analysis.
Such analysis is shown in Fig.\,\ref{fig:comparison2D3D}. 
In this case, all 17 galaxies
from the sample are included in the plot regardless of their error bars:
those finally removed from the analysis with error bars greater than 0.5
are shown in grey, while the coloured symbols correspond to the remaining
8 galaxies. The excellent correspondence
between the measurements from both techniques, even for the 
dismissed galaxies, argues in favor of this analysis
and shows how conservative the uncertainties provided by the PDF are.

\begin{table}
\caption{Intrinsic 3D shape of the different structures in our sample.}
\begin{center}
\resizebox{8cm}{!}{
\begin{tabular}{cccc}
\hline
Galaxy         &  Structure     &        B/A           & C/A            \\
\hline
\hline
NGC\,357         & Bulge          &     0.94$_{0.84}^{0.96}$  &    0.76$_{0.34}^{1.01}$\\
NGC\,357         & Outer bar      &     0.46$_{0.39}^{0.51}$  &    1.96$_{0.69}^{2.00}$\\
NGC\,357         & Inner bar      &     0.91$_{0.74}^{1.00}$  &    0.96$_{0.36}^{1.26}$\\
NGC\,718         & Bulge          &     0.81$_{0.69}^{0.86}$  &    1.21$_{0.34}^{1.89}$\\
NGC\,718         & Outer bar      &     0.36$_{0.16}^{0.41}$  &    1.04$_{0.24}^{1.79}$\\
NGC\,718         & Inner bar      &     0.56$_{0.24}^{0.69}$  &    0.99$_{0.24}^{1.84}$\\
NGC\,2642        & Bulge          &     0.84$_{0.69}^{0.89}$  &    0.91$_{0.26}^{2.00}$\\
NGC\,2642        & Outer bar      &     0.16$_{0.06}^{0.26}$  &    0.69$_{0.16}^{1.24}$\\
NGC\,2642        & Inner bar      &     0.14$_{0.04}^{0.21}$  &    0.34$_{0.11}^{0.79}$\\
NGC\,2681        & Bulge          &     0.91$_{0.79}^{0.94}$  &    0.56$_{0.09}^{1.09}$\\
NGC\,2681        & Outer bar      &     0.76$_{0.69}^{0.76}$  &    1.66$_{0.89}^{2.00}$\\
NGC\,2681        & Inner bar      &     0.89$_{0.76}^{0.91}$  &    1.31$_{0.69}^{1.71}$\\
NGC\,2859        & Bulge          &     0.96$_{0.91}^{1.00}$  &    0.86$_{0.64}^{1.01}$\\
NGC\,2859        & Outer bar      &     0.64$_{0.54}^{0.69}$  &    1.79$_{0.74}^{2.00}$\\
NGC\,2859        & Inner bar      &     0.34$_{0.11}^{0.41}$  &    0.41$_{0.14}^{0.66}$\\
NGC\,2950        & Bulge          &     0.66$_{0.41}^{0.71}$  &    0.59$_{0.19}^{0.81}$\\
NGC\,2950        & Outer bar      &     0.49$_{0.24}^{0.54}$  &    0.36$_{0.11}^{0.64}$\\
NGC\,2950        & Inner bar      &     0.74$_{0.49}^{0.76}$  &    0.36$_{0.09}^{0.69}$\\
NGC\,2962        & Bulge          &     1.00$_{0.96}^{1.00}$  &    0.96$_{0.91}^{0.99}$\\
NGC\,2962        & Outer bar      &     0.76$_{0.44}^{0.79}$  &    0.19$_{0.06}^{0.46}$\\
NGC\,2962        & Inner bar      &     1.00$_{0.96}^{1.00}$  &    0.64$_{0.49}^{0.74}$\\
NGC\,3368        & Bulge          &     0.91$_{0.74}^{0.94}$  &    0.41$_{0.19}^{0.64}$\\
NGC\,3368        & Outer bar      &     0.54$_{0.31}^{0.59}$  &    0.46$_{0.11}^{0.74}$\\
NGC\,3368        & Inner bar      &     0.44$_{0.19}^{0.51}$  &    0.31$_{0.09}^{0.54}$\\
NGC\,3941        & Bulge          &     0.91$_{0.76}^{0.94}$  &    0.61$_{0.39}^{0.79}$\\
NGC\,3941        & Outer bar      &     0.51$_{0.26}^{0.59}$  &    0.29$_{0.09}^{0.51}$\\
NGC\,3941        & Inner bar      &     0.71$_{0.51}^{0.76}$  &    0.56$_{0.16}^{0.81}$\\
NGC\,3945        & Bulge          &     0.94$_{0.86}^{0.96}$  &    0.66$_{0.49}^{0.79}$\\
NGC\,3945        & Outer bar      &     0.81$_{0.66}^{0.86}$  &    1.66$_{1.01}^{1.86}$\\
NGC\,3945        & Inner bar      &     0.51$_{0.21}^{0.64}$  &    0.29$_{0.09}^{0.51}$\\
NGC\,4314        & Bulge          &     0.84$_{0.64}^{0.91}$  &    0.66$_{0.09}^{1.39}$\\
NGC\,4314        & Outer bar      &     0.21$_{0.09}^{0.26}$  &    0.36$_{0.11}^{0.61}$\\
NGC\,4314        & Inner bar      &     0.84$_{0.66}^{0.89}$  &    0.76$_{0.14}^{1.49}$\\
NGC\,4340        & Bulge          &     0.91$_{0.79}^{0.96}$  &    0.54$_{0.34}^{0.74}$\\
NGC\,4340        & Outer bar      &     0.29$_{0.11}^{0.31}$  &    0.51$_{0.14}^{0.84}$\\
NGC\,4340        & Inner bar      &     0.24$_{0.11}^{0.26}$  &    0.61$_{0.16}^{1.01}$\\
NGC\,4503        & Bulge          &     0.84$_{0.64}^{0.91}$  &    0.64$_{0.41}^{0.76}$\\
NGC\,4503        & Outer bar      &     0.61$_{0.31}^{0.66}$  &    0.29$_{0.06}^{0.44}$\\
NGC\,4503        & Inner bar      &     0.44$_{0.24}^{0.51}$  &    0.41$_{0.16}^{0.76}$\\
NGC\,4725        & Bulge          &     0.81$_{0.59}^{0.86}$  &    0.71$_{0.46}^{0.86}$\\
NGC\,4725        & Outer bar      &     0.74$_{0.46}^{0.76}$  &    0.44$_{0.19}^{0.59}$\\
NGC\,4725        & Inner bar      &     0.21$_{0.11}^{0.49}$  &    0.79$_{0.31}^{2.00}$\\
NGC\,5850        & Bulge          &     1.00$_{1.00}^{1.00}$  &    0.94$_{0.91}^{0.99}$\\
NGC\,5850        & Outer bar      &     0.24$_{0.09}^{0.31}$  &    0.24$_{0.09}^{0.36}$\\
NGC\,5850        & Inner bar      &     0.39$_{0.21}^{0.44}$  &    1.86$_{0.61}^{2.00}$\\
NGC\,7280        & Bulge          &     0.89$_{0.64}^{0.96}$  &    0.31$_{0.11}^{0.64}$\\
NGC\,7280        & Outer bar      &     0.61$_{0.34}^{0.69}$  &    0.31$_{0.09}^{0.59}$\\
NGC\,7280        & Inner bar      &     0.11$_{0.04}^{0.16}$  &    0.11$_{0.04}^{0.24}$\\
NGC\,7716        & Bulge          &     0.64$_{0.29}^{0.74}$  &    0.49$_{0.11}^{0.94}$\\
NGC\,7716        & Outer bar      &     0.46$_{0.21}^{0.54}$  &    0.46$_{0.14}^{0.79}$\\
NGC\,7716        & Inner bar      &     0.91$_{0.79}^{0.94}$  &    0.49$_{0.11}^{0.91}$\\
\hline
\end{tabular}
}
\end{center}
\label{tab:intrinsic}
\end{table}

\section{Influence of inner bars in bulge parameters}\label{sec:influence}

\begin{figure*}
 \vspace{2pt}
 \includegraphics[bb= 54 0 620 594, angle=0., width=0.8\textwidth]{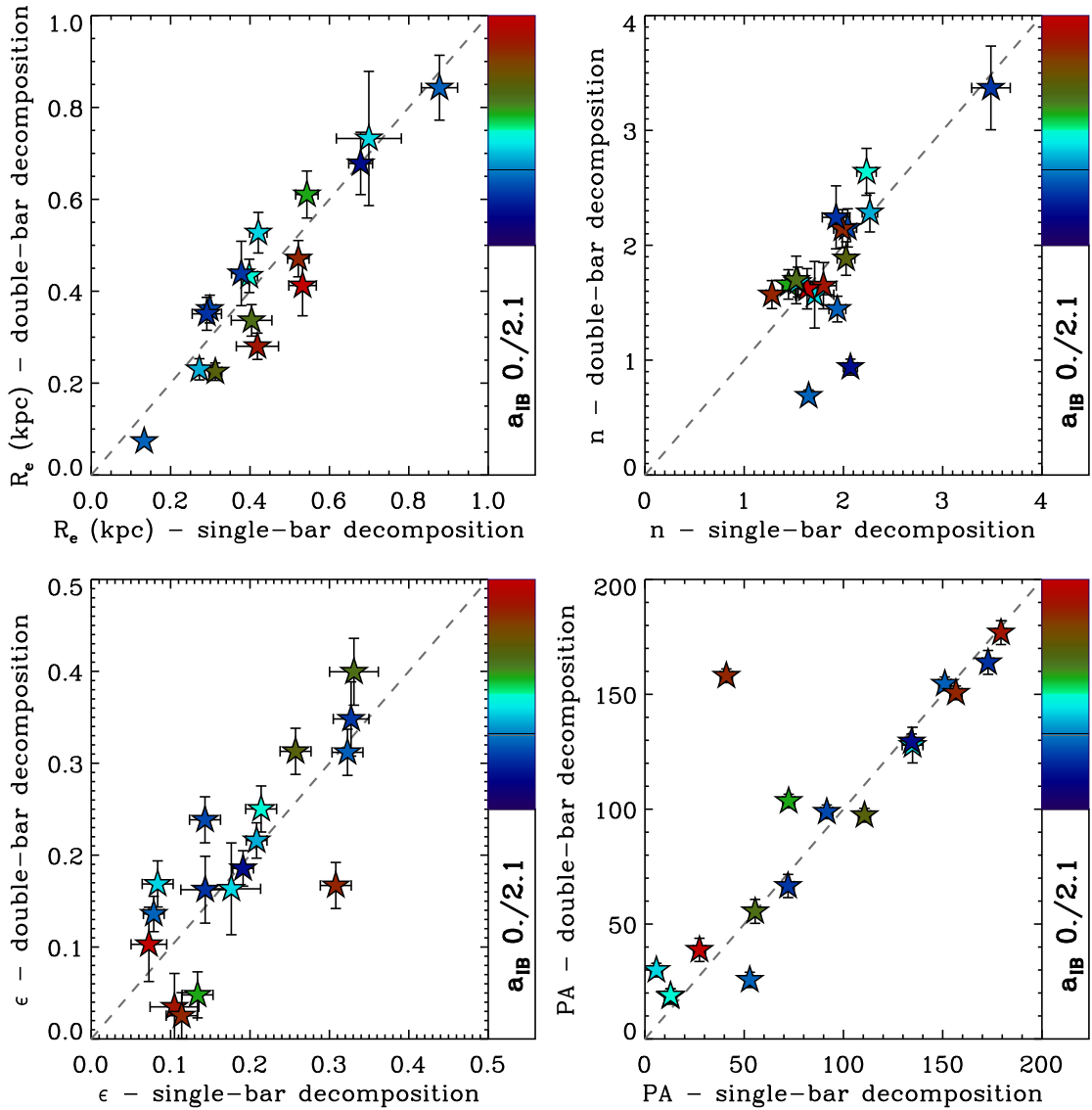}
 \caption{Bulge structural parameters as measured from the photometric decompositions in
$r'$-band including two bars 
(vertical axes) and the most common method of considering only one single bar (horizontal axes). Top left: 
bulge effective radii; top right: \Sersic indices; bottom left: ellipticities; bottom
right: position angles. The grey dashed line marks the 1:1 relationship in each panel.
Values are colour-coded by length of the inner bar as obtained from the double-barred
fitting and indicated by the right-hand side colour bar. 
Except for the \Sersic indices of the most extended
bulges ($n<$1) and the ellipticities of few galaxies with long inner bars, 
no major bias is introduced by disregarding inner bars.}
\label{fig:comparisonbulges}
\end{figure*}

Photometric decompositions have been widely used in the literature
for studying the central regions of galaxies and, particularly, the
nature of bulges \citep[e.g.,][]{Gadotti2009}. 
Current estimates establish that
$\sim$20\% of disc galaxies are double-barred \citep[e.g.,][]{ErwinandSparke2002,Laineetal2002, Erwin2011},
and this is most likely
a lower limit due to the small sizes of inner bars 
($\sim$23\% of the outer bar size as shown in Paper II). Despite their frequent presence,
inner bars have never been taken into account when decomposing galaxies and
not including bars in the photometric decompositions
may significantly affect the derived bulge parameters \citep[e.g.,][]{Laurikainenetal2006,MendezAbreuetal2017},
hampering the conclusions obtained in many works.
In Fig.\,\ref{fig:comparisonbulges} we compare the double- and single-barred
$r'$-band fits of the double-barred galaxies in order to quantify how much bulge measurements
(effective radius, \Sersic index, ellipticity, and position angle)
are affected when inner bars are not included in the analysis.

The effective radii measurements are in good agreement, with a mean difference of 0.07\,kpc
between the single- and double-bars fits.
This value is lower than the typical errors introduced by the adopted methodology
 (see Tables \ref{tab:paramsone1} and \ref{tab:paramsone3}) for
all the galaxies, and therefore negligible.
The agreement
between \Sersic indices
is  also remarkably good except for the two galaxies with $n<$1, one of which
hosts a extremely small bulge (NGC\,2681).
For the remaining galaxies we obtain a mean difference of 0.2, slightly 
higher than the errors from the decompositions. 
The ellipticity shows the largest discrepancy among all parameters, with a
mean difference of 0.05, i.e., larger than 10\% of the typical values. We note
however that such measurement is of the order or only slightly larger than the typical
errors derived from \gasp.
Finally, the position angle is the best behaved measurement, with only one discrepant galaxy
corresponding to a very spherical bulge ($\epsilon=$0.03), where the position angle
is irrelevant.

We can therefore conclude that the bulge parameters
are rather insensitive to disregarding the inner bar in the fits. However, small
differences are observed. With the aim of identifying the main contributor to such
discrepancies, measurements in Fig.\,\ref{fig:comparisonbulges} 
are colour-coded attending to the length of the (dismissed) inner bars. 
A subtle trend pointing at largest inner bars causing the largest differences
in the derived ellipticity and effective radius of the bulges is observed, 
although inner bar size does not account for discrepancies in the \Sersic indices.
The same test has been performed by colour-coding the values with other quantities 
such as \Sersic index, bulge effective radius, and R$_{\rm e}$/a$_{\rm IB}$ ratio. 
No clear correlations have been found, apart from the expected fact that galaxies with
large inner bars with respect to the bulge size (i.e., small R$_{\rm e}$/a$_{\rm IB}$ ratios)
tend to compensate the dismissed inner bar by increasing the size of the
bulge (R$_{\rm e}$).
We therefore conclude that most likely a combination 
of all those parameters (inner bar parameters with respect to bulge parameters)
is responsible for the differences found between the single- and double-barred fits. We
emphasise again that deviations are small and generally within the error bars. 

We note that the inner bar contribution 
to the total galaxy light can be as low as 0.5\% with a mean value
of 4\%, while the outer bar accounts for
[4\%, 28\%] of the total light. It is therefore reasonable to find that bulge 
measurements do not vary in a significant
way when inner bars are not accounted for and therefore their light is included as bulge light.

For the sake of completeness, we have also investigated whether dismissing the inner
bar has any effect over the outer bar parameters. While the length and position angle
of the outer bar are absolutely independent from including the inner bar or not, 
very subtle differences are found for the outer bar ellipticity. However, such differences
do not depend on the inner bar size and they are well within the error bars, thus 
supporting the idea that outer bar parameters are not affected by the inner bar.

\section[]{The nature of bulges within double-barred galaxies}\label{sec:bulgenature}

\subsection[]{Traditional photometric diagnostics}\label{sub:traditional}

\begin{figure}
 \vspace{2pt}
 \includegraphics[angle=0., width=0.45\textwidth]{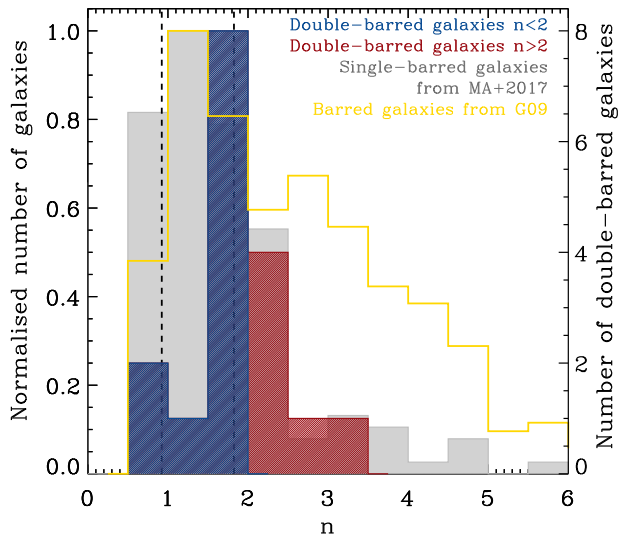}
 \caption{Distribution of the bulge \Sersic indices for our double-barred sample as indicated by the right-hand
side vertical axis. The usual separation $n<$2 and $n>$2 is highlighted in blue
and red, respectively. The grey histogram shows the normalised \Sersic distribution for the bulges of barred galaxies in the CALIFA survey,
as published in \citet{MendezAbreuetal2017}. This sample contains two double-barred galaxies, whose \Sersic values are
indicated with either dashed black lines. The normalised distribution of \Sersic indices for the sample of barred galaxies
of \citet{Gadotti2009} is outlined in yellow. The \citet{MendezAbreuetal2017} and \citet{Gadotti2009} distributions
have been normalised to a maximum value of 1 for the sake of comparison with the sample of double-barred galaxies,
as indicated by the left-hand side vertical axis.}
\label{fig:distributionsersic}
\end{figure}

\begin{figure}
 \vspace{2pt}
 \includegraphics[angle=0., width=0.45\textwidth]{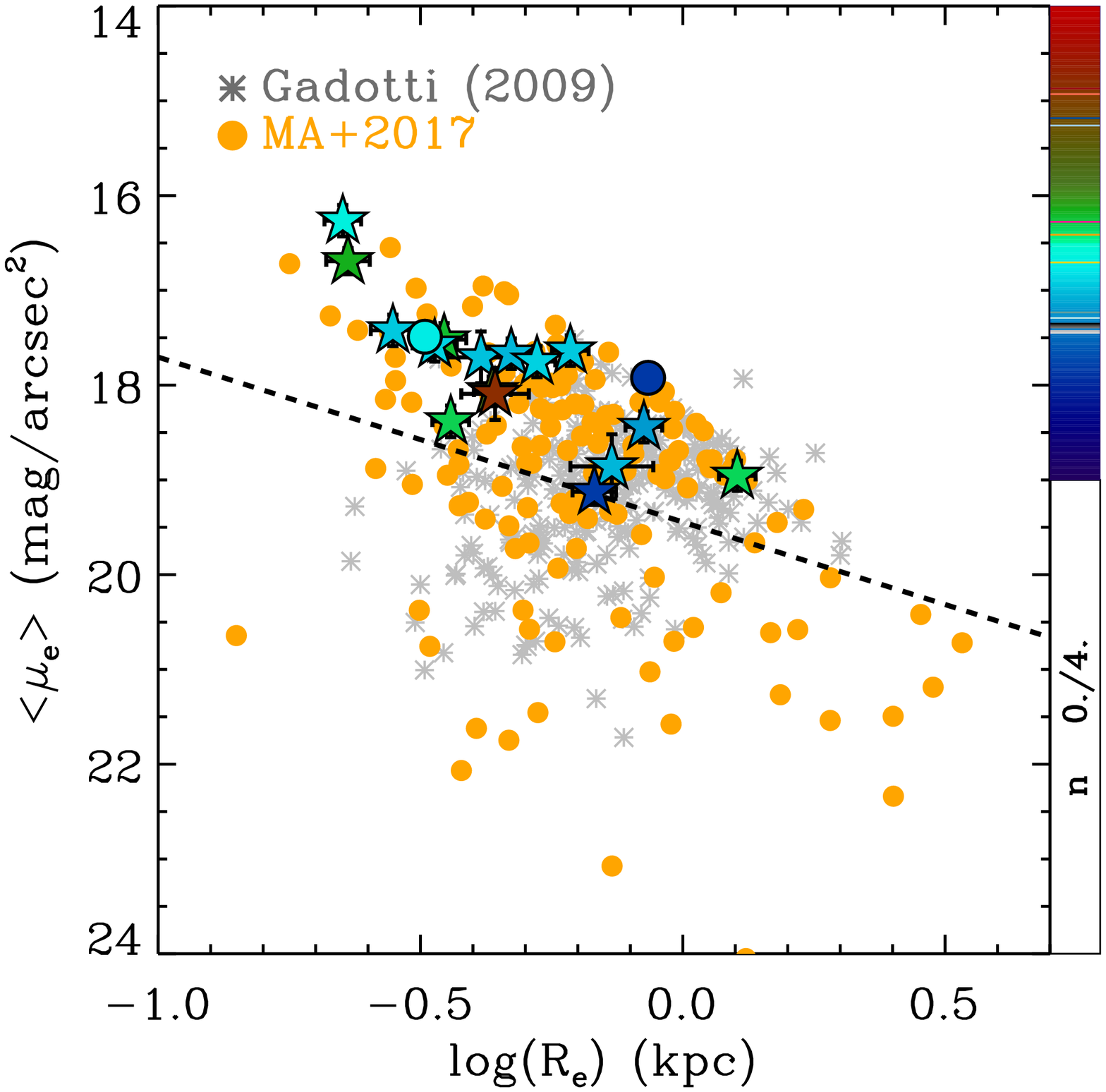}
 \caption{The \citet{Kormendy1977} relation for the bulges of our double-barred sample (stars)
and the two double-barred galaxies included in the analysis of \citet[][circles]{MendezAbreuetal2017}, 
colour-coded by their \Sersic index following the right-hand side colour bar. The corresponding
values for the barred samples of \citet[][orange circles]{MendezAbreuetal2017} and 
\citet[][grey asterisks]{Gadotti2009} are shown for comparison, 
and the dashed black line marks the demarcation line defined
by \citet{Gadotti2009} to separate classical and disc-like bulges. 
The bulges in double-barred galaxies, as well as most
of the bulges in the barred galaxies from \citet{MendezAbreuetal2017}, lay within the classical regime.}
\label{fig:kormendyrelation}
\end{figure}

Figure \ref{fig:distributionsersic} shows the distribution of \Sersic indices for 
the bulges of our double-barred sample. Following the traditional demarcation $n$$=$2,
we find that 6 out of the 17 double-barred galaxies host bulges with $n>$2 and they can 
therefore be considered classical bulges. The remaining 11 galaxies, i.e. the majority
(65\%) of the sample, host disc-like bulges attending to this pure \Sersic index diagnostics.
Since our conclusions might be hampered by the limited size of the sample, we compare
with the 162 single-barred galaxies 
of \citet{MendezAbreuetal2017}, also shown in Fig.\,\ref{fig:distributionsersic}. 
This comparison sample does show a higher incidence of bulges with $n<$2
(77\%), thus
complying with the expectation that single/double-barred galaxies should show 
a large fraction of disc-like bulges. In particular, the two double-barred galaxies in 
\citet{MendezAbreuetal2017} host bulges with $n<$2. For the sake of completeness, Fig.\,\ref{fig:distributionsersic}
shows the results for the 287 barred galaxies included in the sample of \citet{Gadotti2009} as well. 
We remind the reader that this represents a more distant sample where bars are modelled with a \Sersic
profile instead of a Ferrers profile; these differences may account for the higher abundance
of bulges with $n>$2 found in this case (59\%).

The \citet{Kormendy1977} relation is drawn in Fig.\,\ref{fig:kormendyrelation}. As in the previous plot, the 
samples of bulges within barred galaxies from \citet{Gadotti2009}
and \citet{MendezAbreuetal2017} are also shown. The dashed line 
represents the demarcation found by \citet{Gadotti2009} to separate between
the classical and disc-like nature of the bulges. We remark that the CALIFA barred galaxies 
in \citet{MendezAbreuetal2017} are analysed with \gasp; many bulges
in this sample are 
compatible with being classical but there is also a less populated cloud of disc-like-compatible bulges
which extends up to the regime of large bulges. Surprisingly, all bulges in double-barred galaxies lay
at the top sequence of classical bulges regardless of their \Sersic indices.
In fact, no trend with $n$ shows up in Fig.\,\ref{fig:kormendyrelation}.
We must note that the well-known dependence of the Kormendy 
relation with the stellar mass \citep{NigocheNetroetal2008} makes this diagnostic 
not fully reliable. Such problem is shown in \citet{Costantinetal2017},
where it is also claimed that a single line division 
might not be the correct way to separate classical and disc-like bulges 
using this diagnostic.

In summary, the traditional \Sersic index and Kormendy-relation discriminators 
provide contradictory results, with a majority of disc-like bulges in the first case and
all classical bulges in the second case for the double-barred galaxies under study.

\subsection{3D intrinsic shape of bulges}\label{sec:intrinsicbulges}
\begin{figure*}
 \vspace{2pt}
 \includegraphics[angle=0., width=0.8\textwidth]{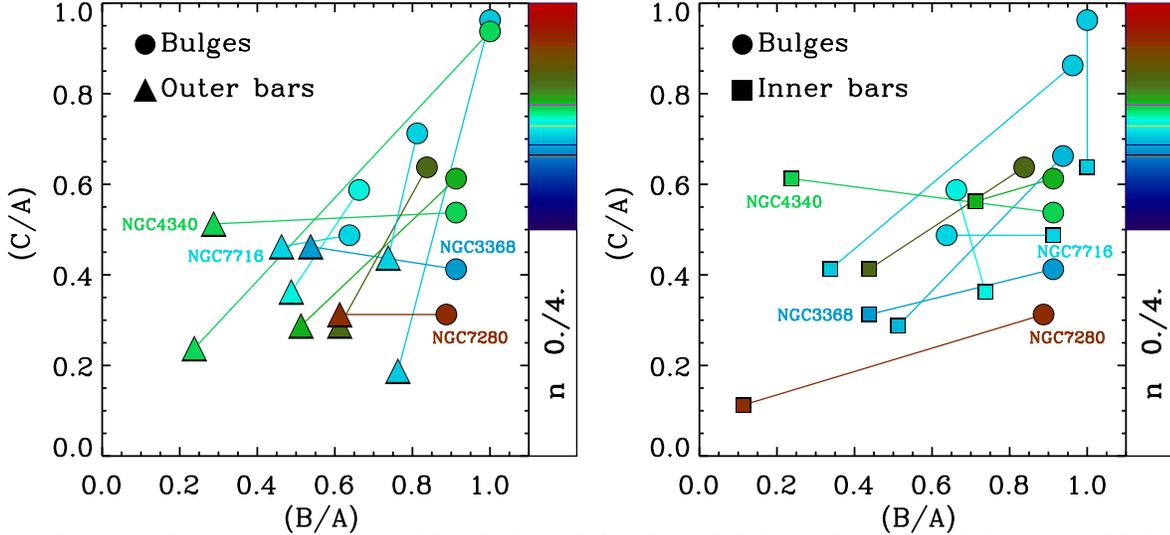}
 \caption{Deprojected semi-major axis ratios of the 3D ellipsoids describing the bulges and outer bars (left panel), and 
bulges and inner bars (right panel). The horizontal axes show ratios in the galaxy plane, while the vertical
axes correspond to the off-plane ratios. We only include galaxies for which the uncertainties in every axis ratio involved
in the plot are $<$0.5. The values for each galaxy are colour-coded by bulge \Sersic index
as indicated in the right-hand side colour bar. The four galaxies discussed in Sect.\,\ref{sec:intrinsicbulges} are identified in both plots. For the sake of clarity, error bars are not shown.}
\label{fig:bulges3D}
\end{figure*}
In \citet{Costantinetal2018b}, different diagnostics to constrain the nature of bulges are put at test.
The main conclusion of that work is that the traditional pure-photometric methods based on projected quantities such
as the \Sersic index do not provide clear separations between classical and disc-like bulges
\citep[see also][]{MendezAbreuetal2018}.
The probabilistic retrieval of the 3D intrinsic shape of bulges represents a better suited 
diagnostic, as it allows the study of the intrinsic flattening of these structures, once isolated
from the remaining components of the galaxy.

Disc-like bulges are expected to be flattened structures with a close-to-circular projection in the galaxy
plane, due to their rotating nature.
Only two bulges in our sample comply with these requirements, as 
indicated by the results shown in Table\,\ref{tab:intrinsic}: NGC\,3368
($B/A=$0.91 and $C/A=$0.41)
and NGC\,7280 ($B/A=$0.89 and $C/A=$0.31). They therefore represent the best candidates to host
disc-like bulges. Note that NGC\,3368 does show a bulge \Sersic index $n<$2, while 
NGC\,7280 would be classified as a classical-bulge host according to the \Sersic index discriminator.

\citet{MendezAbreuetal2018b} discuss why bulges which are intrinsically
flatter than their coexistent single bars represent even stronger candidates to have a disc-like nature.
Such conclusion is based on the fact that the vertical extension of bars is closely related
to that of galaxy discs, so if bulges are as flats as bars, they are also at least as flat as discs,
a result which is hard to reconcile with a classical nature.
The left panel of Fig.\,\ref{fig:bulges3D} shows the intrinsic semiaxis ratios of bulges and outer
bars. Only the 10 out of 17 double-barred galaxies with uncertainties lower than 0.5
in the derived 3D parameters of bulges and outer bars are plotted. Four galaxies show close values of the flattening
between bulges and outer bars. These are NGC\,3368, NGC\,4340, NGC\,7280, and NGC\,7716, being 
NGC\,3368 the only one with an intrinsic flattening for the bulge ($C/A=$0.41) estrictly lower than
that for the outer bar ($C/A=$0.46).
NGC\,3368 and NGC\,7280 therefore remain as the two best candidates for hosting disc-like bulges.
Note that the bulge of NGC\,7716 has an axis ratio  $B/A=$0.64, which does not fulfill the axisymmetric projection requirement.

For the sake of completeness, the right panel of Fig.\,\ref{fig:bulges3D} shows similar results 
but comparing the intrinsic flattening of bulges
and inner bars. A different set of 10 galaxies with low uncertainties affecting bulge and inner
bar measurements are plotted in this case. 
NGC\,3368 and NGC\,7280 are included in this right panel: they host thicker bulges than the
corresponding inner bars, while NGC\,4340 hosts a flatter and almost circular bulge,
and NGC\,3941 and NGC\,7716 show similar intrinsic flattenings for bulges and inner bars.
The intrinsic shape of inner bars has never been studied in the literature and no relationship
between it and the nature of bulges is \emph{a-priori} expected.
A detailed comparison between the 3D shapes of inner and outer bars is presented in Paper II.

\section[]{Discussion}\label{sec:discussion}
\begin{table}
 \centering
  \caption{Bulge colours (within one bulge effective radius), bulge masses, and galaxy masses.}
  \label{tab:colours}
\resizebox{8cm}{!}{ 
 \begin{tabular}{lcccc}
  \hline
  Galaxy  & (g'-i')  & (r'-i') & log(M$_{\rm\star,bulge}$\,(M$_{\odot}$)) & log(M$_{\rm gal}$\,(M$_{\odot}$))\\
\hline
\hline
NGC\,357  & 1.52  & 0.49 & 10.27 & 10.88 \\
NGC\,718  & 1.07  & 0.34 & 9.83  & 10.33 \\
NGC\,2642 & 1.53  & 0.46 & 10.45 & 10.62 \\
NGC\,2681 & -0.04 & -0.10& 7.91  & 9.51  \\
NGC\,2859 & 1.20  & 0.41 & 10.40 & 10.64  \\
NGC\,2950 & 1.18  & 0.36 & 9.97  & 10.38  \\
NGC\,2962 & 1.46  & 0.47 & 10.00 & 10.61  \\
NGC\,3368 & 1.47  & 0.50 & 10.44 & **      \\
NGC\,3941 & 1.20  & 0.39 & 9.80  & 10.41  \\
NGC\,3945 & 1.25  & 0.38 & 9.32  & 10.78  \\
NGC\,4314 & 1.10  & 0.39 & 9.76  & 10.19  \\ 
NGC\,4340 & 1.14  & 0.40 & 9.38  & 9.99   \\
NGC\,4503 & 1.29  & 0.43 & 9.92  & 10.47  \\
NGC\,4725 & 1.28  & 0.49 & 10.24 & ** \\
NGC\,5850 & 1.38  & 0.47 & 10.66 & 11.02  \\
NGC\,7280 & 1.40  & 0.34 & 10.10 & 10.34  \\
NGC\,7716 & 1.38  & 0.46 & 10.53 & 10.67  \\
\hline
\end{tabular}
}
\begin{minipage}{8cm}
** Absolute magnitudes for the galaxies are obtained from SDSS. No values
are provided for these galaxies and therefore masses cannot be computed.
\end{minipage}
\end{table}

We discuss here our results within the context of the two major 
questions brought up in Sect.\,\ref{sec:intro} about bulges within double-barred systems: 
i) do double-barred galaxies host 
a larger fraction of disc-like bulges than non-barred galaxies due to the efficient secular evolution that is expected to take place
in them?; and ii) are bars within bars formed due to the presence of a disc-like component
resulting from secular evolution due to the large-scale bar?\\

Following both the Kormendy relation (Fig.\,\ref{fig:kormendyrelation})
and the 3D intrinsic shape (Fig.\,\ref{fig:bulges3D}) as diagnostics for the bulge nature,
our results agree with double-barred galaxies mostly hosting classical bulges. 
Only the bulge \Sersic indices point towards a larger fraction of disc-like bulges.

Two galaxies among the sample of 17 individuals, NGC\,3368 and NGC\,7280, stand out
as the best candidates for hosting disc-like bulges as suggested by our preferred
diagnostics of intrinsic flattening.
With the aim of understanding better the origin of bulges within double-barred
galaxies, we compute colours for the isolated bulges by using the results from
the 2D photometric decompositions in the different bandpasses (Table\,\ref{tab:colours}). Results among
galaxies can be directly compared as we use colours integrated within one bulge
effective radius, so no bias due to different galaxy sizes affects the results.
All our bulges show rather red colours, with median values of ($g'-i'$)$=$1.28
and ($r'-i'$)$=$0.41. Note these values agree with the results for classical bulges
in, e.g., \citet[][although take into account they integrate inside a fibre instead of 
bulge effective radius]{Gadotti2009}. If disc-like bulges are secularly formed after
a star-forming process triggered by gas which has inflowed along the bar to the central regions,
they should show bluer colours due to a younger stellar content with respect to the older
surrounding structures. NGC\,3368 and NGC\,7280 show colours which are even redder than the median,
with ($g'-i'$) values of 1.47 and 1.40, respectively. Bulge colours do indeed point towards 
an old, classical nature for all bulges in the sample.

Figure\,3 in \citet{FisherandDrory2011} shows that disc-like bulges are more frequent in less
massive galaxies, and therefore any mass bias in our sample needs to be considered as well
\citep[see also Fig.\,12 in][]{Gadotti2009}.
We calculate the galaxy masses by using the colours and the recipe given by \citet{Zibettietal2009}.
Results are shown in Table\,\ref{tab:colours}.
Our galaxies span a wide range of masses between 3$\times$10$^9$ and 1$\times$10$^{11}$\,M$_{\odot}$,
with a median value of 2.6$\times$10$^{10}$\,M$_{\odot}$. Actually, a larger fraction
of disc-like bulges is expected in galaxies at these masses, so no mass bias appears to be affecting
our analysis.\\

In summary, most of the diagnostics indicate our double-barred galaxies host classical bulges.
At
first glance, this finding may look like
against the expectations: secular processes should have played a major role in these
kind of galaxies and therefore a significant presence of disc-like bulges is expected. 
However, we must note here that the traditional
paradigm of a single, either classical or disc-like, bulge in a galaxy 
has recently been challenged by many authors:
the coexistence of several bulges of different nature 
has been observationally demonstrated
by \citet{MendezAbreuetal2014} and \citet{Erwinetal2015}. 
An exhaustive analysis combining photometric with
kinematic information is claimed to be required in order to clearly distinguish the nature and possible 
presence of more than one bulge within a galaxy. If our double-barred galaxies were
hosting composite bulges, the analysis presented here would correspond to the dominant structure
in light, which would be the one revealed in the images and the photometric decompositions.
A previously-formed classical bulge could therefore be hiding the presence of a 
secularly-formed disc-like bulge in our galaxies.

In de Lorenzo-C\'aceres et al. (2018a, submitted), we investigate the kinematics and stellar populations
of two double-barred galaxies for which MUSE TIMER integral-field spectroscopic data is available:
NGC\,1291 and NGC\,5850 (also included in this sample).
For those galaxies we perform 2D photometric decompositions of 3.6\,$\mu$m images including a bulge,
two bars, (truncated) disc and also an inner disc. Whether we can call these inner discs as 
disc-like bulges (and therefore composite bulges as both galaxies do host 
an additional bulge component as well) is a matter of semantics out of the scope of this discussion.
However, we do find a strong connection between the size of the inner discs and the inner bars,
suggestive of a dynamical origin of those inner bars from instabilities in the small-scale discs.
The goal of the photometric study of double-barred galaxies presented here is to study the nature
of the dominant bulges, as we do not have the support of high resolution spectroscopic data 
which may reveal the presence of rotating inner discs. Therefore we cannot discard
the possibility of having a faint inner disc coexisting with the classical bulges in several
(if not all) double-barred galaxies, as it is indeed the case for NGC\,5850
\citep[and also NGC\,357, NGC\,4725 and NGC\,2859, see the kinematic evidences in][ \citealt{deLorenzoCaceresetal2012}, and \citealt{deLorenzoCaceresetal2013}]{deLorenzoCaceresetal2008}.

The B/P structures developed at the inner regions of large stellar bars are considered also as
'bulges' by several authors, attending to the fact that they are central components
which cause an excess of light with respect to the pure disc and bar. Although
we emphasise that B/P are not isolated structures but they rather belong to the bars, we have
performed here a visual inspection of our double-barred galaxies with the aim at looking
for signatures of the presence of a  B/P component coexisting with the dominant bulge. 
In particular, we search for a barlens: an oval component
seen in moderately inclined galaxies which is supposed to be the face-on
counterpart of the X-shaped B/P structure seen edge-on 
\citep{Athanassoulaetal2015,LaurikainenandSalo2017}. Other indicators of the presence
of a B/P are isophotal twists in the innermost bar regions, which tend to show boxy isophotes
accompanied of narrow outermost isophotes 
\citep[the so-called spurs, see][]{ErwinandDebattista2013,ErwinandDebattista2017}.
Note that the bulge component found in our analysis and studied throughout this paper does not correspond to the 
possible B/P revealed by this visual inspection. 7 out of 17 galaxies do not show 
signatures of the presence of a B/P (NGC\,718, NGC\,2681, NGC\,2962, NGC\,3368, NGC\,4503, NGC\,7280, and NGC\,7716), 
while 4 out of 17 galaxies host clear barlenses (NGC\,2950, NGC\,3945, NGC\,4314, NGC\,4340).
For the remaining 6 galaxies, the classification of B/P structures is unclear: NGC\,357, NGC\,2642, 
NGC\,2859, NGC\,3941, and NGC\,4725 show weak elliptical structures, spurs, or isophotal twists that may
be due to B/P. Finally, NGC\,5850 shows a component resembling a barlens but the kinematic study 
presented in de Lorenzo-C\'aceres et al. (2018a, submitted) reveals that it is indeed an inner disc.\\

In order to constrain whether 
there is any further relation between our dominant bulges and the double bars,
correlations between the bar parameters (ellipticity and length) and bulge parameters
(\Sersic index and effective radius) have been searched for, with no particular results found.
As a final note, the fact that all our galaxies are better fitted by including a bulge component, i.e,
all double-barred galaxies do host a bulge, might be relevant for understanding
their formation. Our sample includes some late Hubble
types up to Sbc but no bulgeless galaxy is found. This may point towards a connection
between the presence of inner bars and formation of bulges, and it even poses the question of whether 
the presence of a classical bulge is required for forming an inner bar through, e.g., 
dynamical stabilization of the central regions.

\section[]{Conclusions}\label{sec:conclusions}
We present a complete photometric study of a sample of 17 double-barred galaxies
consisting of two analyses which are for the first time applied to these
kind of objects: i) 2D photometric decompositions including a bulge, inner bar,
outer bar, and (truncated) disc, and ii) 3D deprojections of bulges, inner bars, 
and outer bars thus retrieving their
actual intrinsic shape.
While the photometric properties of bars are explored in the companion Paper II, 
the current work focuses on the properties of bulges. In particular, we 
 constrain the classical vs. disc-like nature of bulges in double-barred
systems. Our galaxy sample spans a wide range in galaxy masses.
The main results are:

\begin{itemize}
\item All double-barred galaxies under study host a bulge component.
\item The bulge properties derived through photometric decompositions 
are not significantly affected by dismissing the presence of the inner bar.
\item 65\% of the galaxies host bulges with \Sersic index $n<$2.
\item All bulges lay at the top sequence of the \citet{Kormendy79} relation,
in the region supposedly populated by classical bulges.
\item No correlations are found between inner bar properties 
(length and ellipticity) and bulge properties (\Sersic index and effective radius).
\item Only 2 out of 17 bulges show an intrinsic shape compatible with 
a disc-like nature, i.e., almost circular in-plane projection and 
flattened off-plane profile: NGC\,3368 and NGC\,7280.
\item 3 out of 17 bulges (including NGC\,3368 and NGC\,7280)
are flatter than their corresponding outer bars. This has been argued to be an indication
of their disc-like nature.
\item Inner bars are either flatter or with similar flattening than their coexisting bulges.
\item All bulges show rather red colours.
\end{itemize}

Most previous results support a classical nature for bulges within double-barred galaxies.
A major incidence of secularly-formed structures such as disc-like bulges
is expected in these galaxies, where two non-axisymmetric structures may help to 
transport gas to the central regions. We note that composite bulges are not studied
in this work and therefore we cannot rule out the possibility of a faint
disc-like component, or even other kind of secularly-formed substructure, 
coexisting within these galaxies; we refer to the 
dominant bulge in this analysis. The presence of a dominant classical bulge in all double-barred galaxies
under study suggests that hosting a hot component may be necessary for the dynamical
development of an inner bar within a barred galaxy.

\section*{Acknowledgments}
We are very grateful to the anonymous referee, whose comments helped improving the 
content and presentation of this paper.
AdLC acknowledges support from grants ST/J001651/1 (UK Science and Technology
Facilities Council - STFC) and AYA2016-77237-C3-1-P (Spanish Ministry of Economy and 
Competitiveness - MINECO).
JMA acknowledges support from the Spanish Ministry of Economy and Competitiveness (MINECO) 
by grant AYA2017-83204-P.

\bibliographystyle{mn2e}
\bibliography{reference}



\appendix

\section[]{\gasp\ structural parameters of 18 double-barred galaxies}\label{sec:appendix}

\begin{onecolumn}
\begin{table}
 \centering
  \caption{ALL PARAMETERS FOR THE DOUBLE-BARRED FITS IN r'-BAND. First set of 9/17 galaxies.}
  \label{tab:paramsr1}
\resizebox{18cm}{!}{
  \begin{tabular}{llccccccccc}
  \hline
& & \multicolumn{1}{c}{NGC\,357} & \multicolumn{1}{c}{NGC\,718} & \multicolumn{1}{c}{NGC\,2642} & NGC\,2681 & NGC\,2859 & NGC\,2950 & NGC\,2962 & NGC\,3368 & NGC\,3941   \\
\hline
\hline
\multirow{6}{*}{Bulge}     & $\mu_e$      &  18.52 $\pm$ 0.26  &  18.42  $\pm$ 0.15  &  19.58  $\pm$ 0.33  &  15.96 $\pm$ 0.13  &  18.52  $\pm$ 0.16  &  16.86  $\pm$ 0.16     &  18.32  $\pm$ 0.15  &  18.91  $\pm$ 0.16  &  17.53  $\pm$ 0.13 \\
                           & R$_e$      &  2.68  $\pm$ 0.43  &  3.15   $\pm$ 0.32  &  2.65   $\pm$ 0.53  &  1.65  $\pm$ 0.17  &  5.63   $\pm$ 0.47  &  2.65   $\pm$ 0.22     &  2.22   $\pm$ 0.23  &  14.59  $\pm$ 1.22  &  3.84   $\pm$ 0.38 \\
                           & n          &  1.62  $\pm$ 0.18  &  2.24   $\pm$ 0.27  &  1.57   $\pm$ 0.29  &  0.69  $\pm$ 0.05  &  1.66   $\pm$ 0.13  &  1.88   $\pm$ 0.15     &  1.65   $\pm$ 0.20  &  1.44   $\pm$ 0.11  &  2.28   $\pm$ 0.17 \\
                           & b/a        &  0.90  $\pm$ 0.04  &  0.84   $\pm$ 0.04  &  0.84   $\pm$ 0.05  &  0.86  $\pm$ 0.02  &  0.95   $\pm$ 0.02  &  0.69   $\pm$ 0.02     &  0.97   $\pm$ 0.04  &  0.69   $\pm$ 0.02  &  0.78   $\pm$ 0.02 \\
                           & PA         &  38.76 $\pm$ 5.05  &  163.89 $\pm$ 5.20  &  127.90 $\pm$ 7.76  &  25.71 $\pm$ 3.31  &  103.53 $\pm$ 2.97  &  97.36  $\pm$ 2.97     &  176.86 $\pm$ 5.20  &  154.56 $\pm$ 2.97  &  18.42  $\pm$ 3.31 \\
                           & B/T        &  0.170             &  0.214              &  0.092              &  0.125             &  0.259              &  0.257                 &  0.135              &  0.214              &  0.224             \\
\hline                                                                                                                                                                                                                                          
\multirow{7}{*}{Disc}      & $\mu_o$      &  20.81 $\pm$ 0.07  &  20.52  $\pm$ 0.07  &  21.11  $\pm$ 0.14  &  20.15 $\pm$ 0.14  &  21.92  $\pm$ 0.14  &  20.17  $\pm$ 0.14     &  21.12  $\pm$ 0.07  &  20.71  $\pm$ 0.07  &  19.29  $\pm$ 0.07 \\
                           & h$_{\rm inner}$ &  29.56 $\pm$ 1.38  &  22.38  $\pm$ 0.85  &  27.61  $\pm$ 1.88  &  33.38 $\pm$ 0.81  &  63.41  $\pm$ 1.05  &  26.33  $\pm$ 0.4  &  33.73  $\pm$ 1.28  &  80.36  $\pm$ 1.33  &  31.17  $\pm$ 0.74    \\ 
                           & b/a        &  0.82  $\pm$ 0.01  &  0.95   $\pm$ 0.01  &  0.97   $\pm$ 0.02  &  0.89  $\pm$ 0.01  &  0.84   $\pm$ 0.01  &  0.64   $\pm$ 0.01     &  0.57   $\pm$ 0.01  &  0.65   $\pm$ 0.01  &  0.65   $\pm$ 0.01 \\
                           & PA         &  18.78 $\pm$ 0.37  &  19.70  $\pm$ 0.58  &  151.41 $\pm$ 0.90  &  12.40 $\pm$ 0.25  &  84.73  $\pm$ 0.45  &  123.80 $\pm$ 0.45     &  2.46   $\pm$ 0.58  &  160.7  $\pm$ 0.45  &  9.15   $\pm$ 0.25 \\
                           & h$_{\rm outer}$ &  13.93 $\pm$ 0.49  &  N/A                &  12.01  $\pm$ 0.42  &  18.70 $\pm$ 0.65  &  N/A                &  N/A              &  N/A                 &  N/A                &  19.99  $\pm$ 0.70     \\
                           & R$_{\rm break}$ &  48.62 $\pm$ 1.65  &  N/A                &  45.55  $\pm$ 1.54  &  88.20 $\pm$ 2.99  &  N/A                &  N/A              &  N/A                 &  N/A                &  39.40  $\pm$ 1.33     \\
                           & D/T        &  0.667             &  0.645              &  0.806              &  0.594             &  0.537              &  0.444                 &  0.587              &  0.527              &  0.614             \\
\hline                                                                                                                                                                                                                                          
\multirow{7}{*}{Outer bar} & $\mu_o$      &  20.73 $\pm$ 0.30  &  20.42  $\pm$ 0.32  &  21.37  $\pm$ 0.45  &  19.43 $\pm$ 0.18  &  20.70  $\pm$ 0.20  &  19.63  $\pm$ 0.20     &  20.69  $\pm$ 0.32  &  20.16  $\pm$ 0.20  &  19.03  $\pm$ 0.18 \\
                           & a          &  38.42 $\pm$ 0.63  &  46.90  $\pm$ 0.78  &  53.08  $\pm$ 1.53  &  39.64 $\pm$ 0.39  &  72.93  $\pm$ 0.79  &  45.59  $\pm$ 0.49     &  49.52  $\pm$ 0.82  &  135.14 $\pm$ 1.46  &  34.98  $\pm$ 0.34 \\
                           & n          &  3                 &  4                  &  3                  &  3                 &  3                  &  3                     &  2                  &  3                  &  2                 \\
                           & b/a        &  0.44  $\pm$ 0.04  &  0.42   $\pm$ 0.02  &  0.24   $\pm$ 0.06  &  0.73  $\pm$ 0.02  &  0.63   $\pm$ 0.03  &  0.53   $\pm$ 0.03     &  0.47   $\pm$ 0.02  &  0.59   $\pm$ 0.03  &  0.44   $\pm$ 0.02 \\
                           & PA         &  119.74 $\pm$ 0.28 &  150.89 $\pm$ 0.33  &  117.11 $\pm$ 0.58  &  83.57 $\pm$ 0.16  &  159.32 $\pm$ 0.13  &  156.05 $\pm$ 0.13     &  174.39 $\pm$ 0.33  &  128.19 $\pm$ 0.13  &  171.29 $\pm$ 0.16 \\
                           & c          &  3                 &  2                  &  4                  &  2                 &  2                  &  2                     &  2                  &  2                  &  2                 \\
                           & Bar/T      &  0.106             &  0.127              &  0.097              &  0.170             &  0.182              &  0.204                 &  0.224              &  0.247              &  0.136             \\ 
\hline                                                                                                                                                                                                                                          
\multirow{7}{*}{Inner bar} & $\mu_o$      &  19.93 $\pm$ 0.98  &  18.50  $\pm$ 0.24  &  18.55  $\pm$ 2.04  &  17.81 $\pm$ 1.09  &  18.33  $\pm$ 1.06  &  18.32  $\pm$ 1.06     &  19.88  $\pm$ 0.24  &  17.55  $\pm$ 1.06  &  19.01  $\pm$ 1.09 \\ 
                           & a          &  13.61 $\pm$ 2.46  &  4.11   $\pm$ 0.70  &  3.26   $\pm$ 1.10  &  13.75 $\pm$ 1.16  &  11.21  $\pm$ 1.15  &  17.18  $\pm$ 1.76     &  14.99  $\pm$ 2.54  &  10.47  $\pm$ 1.07  &  13.42  $\pm$ 1.14 \\
                           & n          &  3                 &  2                  &  2                  &  3                 &  3                  &  4                     &  3                  &  3                  &  3                 \\
                           & b/a        &  0.90  $\pm$ 0.06  &  0.63   $\pm$  0.06 &  0.18   $\pm$ 0.13  &  0.89  $\pm$ 0.03  &  0.37   $\pm$ 0.04  &  0.63   $\pm$ 0.04     &  0.74   $\pm$ 0.06  &  0.44   $\pm$ 0.04  &  0.71   $\pm$ 0.03 \\
                           & PA         &  157.75 $\pm$ 2.85 &  31.99  $\pm$  2.21 &  160.04 $\pm$ 3.64  &  75.77 $\pm$ 2.09  &  62.32  $\pm$ 1.30  &  140.56 $\pm$ 1.30     &  0.96   $\pm$ 2.21  &  133.83 $\pm$ 1.30  &  165.58 $\pm$ 2.09 \\
                           & c          &  2                 &  4                  &  2                  &  2                 &  4                  &  2                     &  2                  &  4                  &  2                 \\
                           & Bar/T      &  0.057             &  0.013              &  0.005              &  0.111             &  0.022              &  0.094                 &  0.053              &  0.012              &  0.026             \\
\hline
\end{tabular}
}
\begin{minipage}{16cm}
Intensities are in magnitudes/arcsec$^2$.\\
Effective radii, disc scalelengths and break radii, and bar lengths are provided in arcsec.\\
Position angles are given in degrees from North to East.\\
b/a is the minor-to-major axis of the projected ellipsoid  (ellipticity is $\epsilon =$1-b/a).\\\\
The analytical functions describing the structures are explained in Sect.\,\ref{sec:GASP2D}.

\end{minipage}
\end{table}
\end{onecolumn}

\begin{onecolumn}
\begin{table}
  \caption{ALL PARAMETERS FOR THE DOUBLE-BARRED FITS IN r'-BAND. Last set of 8/17 galaxies}
  \label{tab:paramsr3}
\resizebox{16.4cm}{!}{
  \begin{tabular}{llccccccccc}
  \hline
& & NGC\,3945 & NGC\,4314 & NGC\,4340 & NGC\,4503 & NGC\,4725 & NGC\,5850 & NGC\,7280 & NGC\,7716 \\
\hline
\hline
\multirow{6}{*}{Bulge}     & $\mu_e$           & 18.39  $\pm$ 0.16   & 19.57  $\pm$ 0.13 & 19.16 $\pm$ 0.16  & 18.93  $\pm$ 0.16  & 18.50  $\pm$ 0.16  & 20.00  $\pm$ 0.16  & 18.93  $\pm$ 0.26  & 17.97 $\pm$ 0.15  \\
                           & R$_e$           & 5.72   $\pm$ 0.48   & 10.94  $\pm$ 1.09 & 6.01  $\pm$ 0.50  & 5.05   $\pm$ 0.42  & 6.80   $\pm$ 0.57  & 7.77   $\pm$ 0.65  & 3.71   $\pm$ 0.59  & 2.04  $\pm$ 0.21  \\
                           & n               & 1.57   $\pm$ 0.12   & 0.94   $\pm$ 0.07 & 2.15  $\pm$ 0.17  & 2.64   $\pm$ 0.20  & 1.68   $\pm$ 0.13  & 2.14   $\pm$ 0.17  & 3.37   $\pm$ 0.36  & 1.70  $\pm$ 0.21  \\
                           & b/a             & 0.83   $\pm$ 0.02   & 0.81   $\pm$ 0.02 & 0.76  $\pm$ 0.02  & 0.75   $\pm$ 0.03  & 0.83   $\pm$ 0.03  & 0.97   $\pm$ 0.03  & 0.65   $\pm$ 0.04  & 0.60  $\pm$ 0.04  \\
                           & PA              & 150.63 $\pm$ 2.97   & 129.41 $\pm$ 3.31 & 98.85 $\pm$ 2.97  & 18.90  $\pm$ 2.97  & 29.97   $\pm$ 2.97  & 158.06 $\pm$ 2.97  & 66.56  $\pm$ 5.05  & 55.52 $\pm$ 5.20  \\
                           & B/T             & 0.258               & 0.214             & 0.250             & 0.205              & 0.091              & 0.182              & 0.280              & 0.144             \\
\hline                                                                                                                                                     
\multirow{7}{*}{Disc}      & $\mu_o$           & 21.20  $\pm$ 0.07   & 21.49  $\pm$ 0.07 & 20.73 $\pm$ 0.07  & 19.45  $\pm$ 0.07  & 20.06  $\pm$ 0.07  & 21.67  $\pm$ 0.07  & 20.59  $\pm$ 0.07  & 20.11 $\pm$ 0.07  \\
                           & h$_{\rm inner}$  & 46.59  $\pm$ 0.77  & 62.51  $\pm$ 1.51 & 34.17 $\pm$ 0.57  & 26.40  $\pm$ 0.44  & 85.43  $\pm$ 1.42  & 61.49  $\pm$ 1.02  & 26.74  $\pm$ 1.25  & 16.34 $\pm$ 0.62  \\ 
                           & b/a             & 0.70   $\pm$ 0.01   & 0.92   $\pm$ 0.01 & 0.67  $\pm$ 0.01  & 0.46   $\pm$ 0.01  & 0.50   $\pm$ 0.01  & 0.75   $\pm$ 0.01  & 0.65   $\pm$ 0.01  & 0.82  $\pm$ 0.01  \\
                           & PA              & 157.45 $\pm$ 0.45   & 135.48 $\pm$ 0.25 & 91.94 $\pm$ 0.45  & 9.83   $\pm$ 0.45  & 43.10  $\pm$ 0.45  & 137.26 $\pm$ 0.45  & 74.19  $\pm$ 0.37  & 41.49 $\pm$ 0.58  \\
                           & h$_{\rm outer}$  & N/A                 & 25.56  $\pm$ 0.89 & N/A               & N/A                & N/A                & N/A                & 11.12  $\pm$ 0.39  & N/A               \\
                           & R$_{\rm break}$  & N/A                 & 106.49 $\pm$ 3.6 & N/A               & N/A                & N/A                & N/A                & 44.65  $\pm$ 1.51  & N/A               \\
                           & D/T             & 0.466               & 0.486             & 0.624             & 0.713              & 0.863              & 0.698              & 0.629              & 0.735             \\
\hline                                                                                                                                                     
\multirow{7}{*}{Outer bar} & $\mu_o$           & 20.27  $\pm$ 0.20   & 20.28  $\pm$ 0.18 & 21.72 $\pm$ 0.20  & 20.41  $\pm$ 0.20  & 20.43  $\pm$ 0.20  & 21.52  $\pm$ 0.20  & 20.48  $\pm$ 0.30  & 21.28 $\pm$ 0.32  \\
                           & a               & 57.00  $\pm$ 0.62   & 126.16 $\pm$ 1.24 & 62.90 $\pm$ 0.68  & 32.60  $\pm$ 0.35  & 59.09 $\pm$ 0.64  & 130.87 $\pm$ 1.42  & 20.24  $\pm$ 0.33  & 30.38 $\pm$ 0.50  \\
                           & n               & 2                   & 3                 & 1                 & 2                  & 3                  & 4                  & 2                  & 2                 \\
                           & b/a             & 0.66   $\pm$ 0.03   & 0.25   $\pm$ 0.02 & 0.43  $\pm$ 0.03  & 0.46   $\pm$ 0.02  & 0.61   $\pm$ 0.03  & 0.25   $\pm$ 0.03  & 0.51   $\pm$ 0.04  & 0.48  $\pm$ 0.02  \\
                           & PA              & 73.54  $\pm$ 0.13   & 146.77 $\pm$ 0.16 & 30.98 $\pm$ 0.13  & 176.76 $\pm$ 0.13  & 35.81  $\pm$ 0.13  & 116.04 $\pm$ 0.13  & 57.25  $\pm$ 0.28  & 17.82 $\pm$ 0.33  \\
                           & c               & 2                   & 2                 & 4                 & 2                  & 3                  & 2                  & 2                  & 3                 \\
                           & Bar/T           & 0.219               & 0.276             & 0.109             & 0.065              & 0.040              & 0.110              & 0.067              & 0.073             \\ 
\hline                                                                                                                                                     
\multirow{7}{*}{Inner bar} & $\mu_o$           & 18.88  $\pm$ 1.06   & 16.95  $\pm$ 1.09 & 18.41 $\pm$ 1.06  & 19.63  $\pm$ 1.06  & 18.28  $\pm$ 1.06  & 19.21  $\pm$ 1.06  & 16.87  $\pm$ 0.98  & 19.27 $\pm$ 0.24  \\ 
                           & a               & 21.90  $\pm$ 2.24   & 4.92   $\pm$ 0.42 & 8.83  $\pm$ 0.90  & 12.03  $\pm$ 1.23  & 11.54  $\pm$ 1.18  & 11.00  $\pm$ 1.13  & 4.00   $\pm$ 0.72  & 8.38  $\pm$ 1.42  \\
                           & n               & 3                   & 4                 & 4                 & 4                  & 3                  & 2                  & 2                  & 3                 \\
                           & b/a             & 0.41   $\pm$ 0.04   & 0.83   $\pm$ 0.03 & 0.36  $\pm$ 0.04  & 0.69   $\pm$ 0.04  & 0.34   $\pm$ 0.04  & 0.25   $\pm$ 0.04  & 0.17   $\pm$ 0.06  & 0.84  $\pm$ 0.06  \\
                           & PA              & 159.97 $\pm$ 1.30   & 158.37 $\pm$ 2.09 & 19.99 $\pm$ 1.30  & 146.14 $\pm$ 1.30  & 137.33   $\pm$ 1.30  & 46.83  $\pm$ 1.30  & 114.24 $\pm$ 2.85  & 56.17 $\pm$ 2.21  \\
                           & c               & 3                   & 2                 & 4                 & 2                  & 5                  & 4                  & 1                 & 2                 \\
                           & Bar/T           & 0.056               & 0.024             & 0.017             & 0.017              & 0.006              & 0.010              & 0.024              & 0.047             \\
\hline
\end{tabular}
}
\end{table}

\begin{table}
 \centering
  \caption{ALL PARAMETERS FOR THE DOUBLE-BARRED FITS IN g'-BAND. First set of 9/17 galaxies.}
  \label{tab:paramsg1}
\resizebox{18cm}{!}{
  \begin{tabular}{llccccccccc}
  \hline
& & \multicolumn{1}{c}{NGC\,357} & \multicolumn{1}{c}{NGC\,718} & \multicolumn{1}{c}{NGC\,2642} & NGC\,2681 & NGC\,2859 & NGC\,2950 & NGC\,2962 & NGC\,3368 & NGC\,3941  \\
\hline
\hline
\multirow{6}{*}{Bulge}     & $\mu_e$          & 19.55 $\pm$ 0.33  & 19.04   $\pm$ 0.31   & 20.42   $\pm$ 0.33  & 16.21 $\pm$ 0.13  & 19.16 $\pm$ 0.15  & 17.74  $\pm$ 0.15 & 19.25  $\pm$ 0.31  & 19.82   $\pm$ 0.16   & 18.44 $\pm$ 0.13   \\
                           & R$_e$          & 2.76  $\pm$ 0.55  & 3.02    $\pm$ 0.53   & 2.62    $\pm$ 0.52  & 1.48  $\pm$ 0.15  & 5.29  $\pm$ 0.54  & 2.77   $\pm$ 0.28 & 2.28   $\pm$ 0.40  & 15.36   $\pm$ 1.29   & 3.97  $\pm$ 0.40   \\
                           & n              & 1.60  $\pm$ 0.29  & 2.02    $\pm$ 0.40   & 1.19    $\pm$ 0.22  & 2.02  $\pm$ 0.15  & 1.50  $\pm$ 0.18  & 1.67   $\pm$ 0.20 & 1.46   $\pm$ 0.29  & 1.37    $\pm$ 0.11   & 2.59  $\pm$ 0.19   \\
                           & b/a            & 0.87  $\pm$ 0.05  & 0.83    $\pm$ 0.05   & 0.78    $\pm$ 0.05  & 0.86  $\pm$ 0.02  & 0.92  $\pm$ 0.04  & 0.71   $\pm$ 0.04 & 0.93   $\pm$ 0.05  & 0.64    $\pm$ 0.03   & 0.78  $\pm$ 0.02   \\
                           & PA             & 33.72 $\pm$ 7.76  & 164.48  $\pm$ 7.79   & 131.33  $\pm$ 7.76  & 29.70 $\pm$ 3.31  & 92.96 $\pm$ 5.20  & 100.99 $\pm$ 5.20 & 7.11   $\pm$ 7.79  & 152.84  $\pm$ 2.97   & 19.79 $\pm$ 3.31   \\
                           & B/T            & 0.164             & 0.203                & 0.053               & 0.166             & 0.204             & 0.244             & 0.110              & 0.173                & 0.214              \\
\hline                                                                                                                                                                                                                                                                                                                                                                                                          
\multirow{7}{*}{Disc}      & $\mu_o$          & 21.76 $\pm$ 0.14  & 21.25   $\pm$ 0.09   & 21.82   $\pm$ 0.14  & 20.47 $\pm$ 0.14  & 22.53 $\pm$ 0.07  & 20.99  $\pm$ 0.07 & 22.12  $\pm$ 0.09  & 21.47   $\pm$ 0.09   & 20.08 $\pm$ 0.14   \\
                           & h$_{\rm inner}$ & 29.70 $\pm$ 2.03  & 22.70   $\pm$ 1.77   & 37.83   $\pm$ 2.58  & 33.94 $\pm$ 0.82  & 72.72 $\pm$ 2.77  & 27.69  $\pm$ 1.05 & 38.96  $\pm$ 3.03  & 84.96   $\pm$ 1.41   & 32.34 $\pm$ 0.78   \\
                           & b/a            & 0.83  $\pm$ 0.02  & 0.94    $\pm$ 0.01   & 0.97    $\pm$ 0.02  & 0.88  $\pm$ 0.01  & 0.76  $\pm$ 0.01  & 0.64   $\pm$ 0.01 & 0.57   $\pm$ 0.01  & 0.67    $\pm$ 0.01   & 0.65  $\pm$ 0.01   \\
                           & PA             & 19.24 $\pm$ 0.90  & 24.29   $\pm$ 0.73   & 155.77  $\pm$ 0.90  & 13.76 $\pm$ 0.25  & 78.24 $\pm$ 0.58  & 124.60 $\pm$ 0.58 & 4.41   $\pm$ 0.73  & 161.45  $\pm$ 0.45   & 10.05 $\pm$ 0.25   \\ 
                           & h$_{\rm outer}$ & 14.38 $\pm$ 0.50  & N/A                  & 14.56   $\pm$ 0.51  & 21.22 $\pm$ 0.74  & N/A               & N/A               & N/A                & N/A                  & 20.89 $\pm$ 0.73   \\
                           & R$_{\rm break}$ & 47.57 $\pm$ 1.61  & N/A                  & 39.77   $\pm$ 1.35  & 87.60 $\pm$ 2.96  & N/A               & N/A               & N/A                & N/A                  & 37.81 $\pm$ 1.28   \\
                           & D/T            & 0.681             & 0.649                & 0.874               & 0.595             & 0.628             & 0.453             & 0.624              & 0.562                & 0.621              \\
\hline                                                                                                                                                                                                                                                                                                                                                                                                            
\multirow{7}{*}{Outer bar} & $\mu_o$          & 21.72 $\pm$ 0.45  & 21.22   $\pm$ 0.42   & 22.32   $\pm$ 0.45  & 19.88 $\pm$ 0.18  & 21.35 $\pm$ 0.32  & 20.38  $\pm$ 0.32 & 21.49  $\pm$ 0.42  & 20.87   $\pm$ 0.20   & 19.76 $\pm$ 0.18   \\
                           & a              & 37.64 $\pm$ 1.08  & 48.77   $\pm$ 1.46   & 54.07   $\pm$ 1.55  & 41.34 $\pm$ 0.41  & 70.69 $\pm$ 1.17  & 46.05  $\pm$ 0.76 & 49.66  $\pm$ 1.48  & 136.42  $\pm$ 1.48   & 34.86 $\pm$ 0.34   \\
                           & n              & 3                 & 4                    & 3                   & 3                 & 3                 & 3                 & 2                  & 3                    & 2                  \\
                           & b/a            & 0.44  $\pm$ 0.06  & 0.43    $\pm$ 0.09   & 0.24    $\pm$ 0.06  & 0.71  $\pm$ 0.02  & 0.60  $\pm$ 0.02  & 0.55   $\pm$ 0.02 & 0.48   $\pm$ 0.09  & 0.61    $\pm$ 0.03   & 0.44  $\pm$ 0.02   \\
                           & PA             & 119.74$\pm$ 0.57  & 150.08  $\pm$ 0.50   & 118.46  $\pm$ 0.57  & 85.55 $\pm$ 0.16  & 160.20$\pm$ 0.33  & 156.86 $\pm$ 0.33 & 174.30 $\pm$ 0.50  & 130.34  $\pm$ 0.13   & 172.29$\pm$ 0.16   \\
                           & c              & 3                 & 2                    & 4                   & 2                 & 2                 & 2                 & 2                  & 2                    & 2                  \\
                           & Bar/T          & 0.099             & 0.130                & 0.067               & 0.154             & 0.154             & 0.211             & 0.217              & 0.254                & 0.136              \\
\hline                                                                                                                                                                                                                                                                                                                                                                                                          
\multirow{7}{*}{Inner bar} & $\mu_o$          & 20.91 $\pm$ 2.04  & 19.02   $\pm$  1.77  & 18.21   $\pm$ 2.04  & 18.39 $\pm$ 1.09  & 19.18 $\pm$ 0.24  & 19.17  $\pm$ 0.24 & 20.72  $\pm$ 1.77  & 18.60   $\pm$ 1.06   & 19.60 $\pm$ 1.09   \\
                           & a              & 13.99 $\pm$ 4.73  & 4.11    $\pm$  1.33  & 2.15    $\pm$ 0.73  & 14.12 $\pm$ 1.20  & 10.71 $\pm$ 1.81  & 17.69  $\pm$ 2.99 & 14.98  $\pm$ 4.84  & 11.74   $\pm$ 1.20   & 13.21 $\pm$ 1.12   \\
                           & n              & 3                 & 2                    & 2                   & 3                 & 3                 & 4                 & 3                  & 3                    & 3                  \\
                           & b/a            & 0.87  $\pm$ 0.13  & 0.69    $\pm$  0.95  & 0.26    $\pm$ 0.13  & 0.85  $\pm$ 0.03  & 0.30  $\pm$ 0.06  & 0.65   $\pm$ 0.06 & 0.75   $\pm$ 0.95  & 0.42    $\pm$ 0.04   & 0.72  $\pm$ 0.03   \\
                           & PA             & 151.62$\pm$ 3.64  & 35.37   $\pm$  0.38  & 166.23  $\pm$ 3.64  & 91.29 $\pm$ 2.09  & 61.67 $\pm$ 2.21  & 142.39 $\pm$ 2.21 & 179.84 $\pm$ 0.38  & 145.63  $\pm$ 1.30   & 168.06$\pm$ 2.09   \\
                           & c              & 2                 & 4                    & 2                   & 2                 & 4                 & 2                 & 2                  & 4                    & 2                  \\
                           & Bar/T          & 0.057             & 0.018                & 0.006               & 0.085             & 0.013             & 0.092             & 0.049              & 0.011                & 0.029              \\

\hline
\end{tabular}
}
\end{table}

\begin{table}
  \caption{ALL PARAMETERS FOR THE DOUBLE-BARRED FITS IN g'-BAND. Last set of 8/17 galaxies}
  \label{tab:paramsg3}
\resizebox{16.4cm}{!}{
  \begin{tabular}{llccccccccc}
  \hline
& & NGC\,3945 & NGC\,4314 & NGC\,4340 & NGC\,4503 & NGC\,4725 & NGC\,5850 & NGC\,7280 & NGC\,7716 \\
\hline
\hline
\multirow{6}{*}{Bulge}     & $\mu_e$          & 20.71   $\pm$ 0.15   & 20.16   $\pm$ 0.26 & 20.05 $\pm$ 0.15  & 19.81 $\pm$ 0.15  & 19.22  $\pm$ 0.16  & 20.85   $\pm$ 0.15  & 19.87 $\pm$ 0.33  & 18.08 $\pm$ 0.31  \\
                           & R$_e$          & 11.50   $\pm$ 1.17   & 10.32   $\pm$ 1.65 & 6.32  $\pm$ 0.64  & 5.14  $\pm$ 0.52  & 6.14   $\pm$ 0.51  & 7.45    $\pm$ 0.76  & 4.26  $\pm$ 0.85  & 1.37  $\pm$ 0.24  \\
                           & n              & 3.41    $\pm$ 0.42   & 0.80    $\pm$ 0.09 & 2.58  $\pm$ 0.32  & 2.60  $\pm$ 0.32  & 1.86   $\pm$ 0.14  & 2.12    $\pm$ 0.26  & 2.16  $\pm$ 0.40  & 1.93  $\pm$ 0.38  \\
                           & b/a            & 0.88    $\pm$ 0.04   & 0.82    $\pm$ 0.04 & 0.75  $\pm$ 0.04  & 0.75  $\pm$ 0.04  & 0.85   $\pm$ 0.03  & 0.97    $\pm$ 0.04  & 0.60  $\pm$ 0.05  & 0.45  $\pm$ 0.05  \\
                           & PA             & 149.56  $\pm$ 5.20   & 123.02  $\pm$ 5.05 & 98.14 $\pm$ 5.20  & 20.03 $\pm$ 5.20  & 34.43   $\pm$ 2.97  & 148.30  $\pm$ 5.20  & 67.54 $\pm$ 7.76  & 53.90 $\pm$ 7.79  \\
                           & B/T            & 0.340                & 0.207              & 0.256             & 0.197             & 0.070              & 0.138               & 0.232             & 0.082             \\
\hline                                                                                                                                                                                                                                                                                                                                                                                                    
\multirow{7}{*}{Disc}      & $\mu_o$          & 22.67   $\pm$ 0.07   & 22.17   $\pm$ 0.07 & 21.52 $\pm$ 0.07  & 20.27 $\pm$ 0.07  & 20.93  $\pm$ 0.07  & 22.37   $\pm$ 0.07  & 21.31 $\pm$ 0.14  & 20.88 $\pm$ 0.09  \\
                           & h$_{\rm inner}$ &69.07    $\pm$ 2.63   & 59.84   $\pm$ 2.79 & 34.96 $\pm$ 1.33  & 26.45 $\pm$ 1.01  & 99.59 $\pm$ 1.65  & 65.99   $\pm$ 2.51  & 26.90 $\pm$ 1.84  & 18.18 $\pm$ 1.42  \\
                           & b/a            & 0.67    $\pm$ 0.01   & 0.92    $\pm$ 0.01 & 0.67  $\pm$ 0.01  & 0.46  $\pm$ 0.01  & 0.50   $\pm$ 0.01  & 0.74    $\pm$ 0.01  & 0.65  $\pm$ 0.02  & 0.82  $\pm$ 0.01  \\
                           & PA             & 166.07  $\pm$ 0.58   & 137.31  $\pm$ 0.37 & 91.68 $\pm$ 0.58  & 9.97  $\pm$ 0.58  & 42.99  $\pm$ 0.45  & 138.64  $\pm$ 0.58  & 74.48 $\pm$ 0.90  & 37.85 $\pm$ 0.73  \\
                           & h$_{\rm outer}$ & N/A                  & 26.52   $\pm$ 0.93 & N/A               & N/A               & N/A                & N/A                 & 10.90 $\pm$ 0.38  & N/A               \\
                           & R$_{\rm break}$ & N/A                  & 108.86  $\pm$ 3.68 & N/A               & N/A               & N/A                & N/A                 & 44.51 $\pm$ 1.51  & N/A               \\
                           & D/T            & 0.455                & 0.500              & 0.620             & 0.721             & 0.891              & 0.762               & 0.644             & 0.781             \\
\hline                                                                                                                                                                                                                                                                                                                                                                                                     
\multirow{7}{*}{Outer bar} & $\mu_o$          & 21.42   $\pm$ 0.32   & 21.04   $\pm$ 0.30 & 22.47 $\pm$ 0.32  & 21.22 $\pm$ 0.32  & 21.00  $\pm$ 0.20  & 22.40   $\pm$ 0.32  & 21.50 $\pm$ 0.45  & 21.92 $\pm$ 0.42  \\
                           & a              & 58.16   $\pm$ 0.96   & 125.89  $\pm$2.05   & 62.42 $\pm$ 1.03  & 32.01 $\pm$ 0.53  & 54.26 $\pm$ 0.59  & 135.53  $\pm$ 2.24  & 21.19 $\pm$ 0.61  & 30.48 $\pm$ 0.91  \\
                           & n              & 2                    & 3                  & 1                 & 2                 & 3                  & 4                   & 2                 & 2                 \\
                           & b/a            &  0.63    $\pm$ 0.02   & 0.26   $\pm$ 0.04 & 0.44  $\pm$ 0.02  & 0.48  $\pm$ 0.02  & 0.66   $\pm$ 0.03  & 0.24    $\pm$ 0.02  & 0.53  $\pm$ 0.06  & 0.48  $\pm$ 0.09  \\
                           & PA             & 74.99   $\pm$ 0.33   & 147.29  $\pm$ 0.28 & 30.81 $\pm$ 0.33  & 176.82$\pm$ 0.33  & 36.25  $\pm$0.13   & 114.98  $\pm$ 0.33  & 55.43 $\pm$ 0.57  & 17.19 $\pm$ 0.50  \\
                           & c              & 2                    & 2                  & 4                 & 2                 & 3                  & 2                   & 2                 & 3                 \\
                           & Bar/T          & 0.138                & 0.275              & 0.108             & 0.065             & 0.035              & 0.091               & 0.060             & 0.071             \\
\hline                                                                                                                                                                                                                                                                                                                            
\multirow{7}{*}{Inner bar} & $\mu_o$          & 19.23   $\pm$ 0.24   & 17.63   $\pm$ 0.98 & 18.79 $\pm$ 0.24  & 20.62 $\pm$ 0.24  & 19.21  $\pm$ 1.06  & 19.68   $\pm$ 0.24  & 16.44 $\pm$ 2.04  & 19.47 $\pm$ 1.77  \\
                           & a              & 19.44   $\pm$ 3.29   & 4.54    $\pm$ 0.82 & 8.56  $\pm$ 1.45  & 13.29 $\pm$ 2.25  & 12.84  $\pm$ 1.32  & 10.52   $\pm$ 1.78  & 2.41  $\pm$ 0.81  & 8.57  $\pm$ 2.77  \\
                           & n              & 3                    & 4                  & 4                 & 4                 & 3                  & 2                   & 2                 & 3                 \\
                           & b/a            & 0.47    $\pm$ 0.06   & 0.69    $\pm$ 0.06 & 0.25  $\pm$ 0.06  &  0.64 $\pm$ 0.06  & 0.26   $\pm$ 0.04  & 0.19    $\pm$ 0.06  & 0.41  $\pm$ 0.13  & 0.76  $\pm$ 0.95  \\
                           & PA             & 161.04  $\pm$ 2.21   & 165.23  $\pm$ 2.85 & 21.76 $\pm$ 2.21  & 148.87$\pm$ 2.21  & 140.93  $\pm$ 1.30  & 45.91   $\pm$ 2.21  & 106.65$\pm$ 3.64  & 56.82 $\pm$ 0.38  \\
                           & c              & 3                    & 2                  & 4                 & 2                 & 5                  & 4                   & 1                 & 2                 \\
                           & Bar/T          & 0.067                & 0.018              & 0.016             & 0.017             & 0.004              & 0.009               & 0.064             & 0.066             \\                   
\hline
\end{tabular}
}
\end{table}

\begin{table}
 \centering
  \caption{ALL PARAMETERS FOR THE DOUBLE-BARRED FITS IN i'-BAND. First set of 9/17 galaxies.}
  \label{tab:paramsi1}
\resizebox{18cm}{!}{
  \begin{tabular}{llccccccccc}
  \hline
& & \multicolumn{1}{c}{NGC\,357} & \multicolumn{1}{c}{NGC\,718} & \multicolumn{1}{c}{NGC\,2642} & NGC\,2681 & NGC\,2859 & NGC\,2950 & NGC\,2962 & NGC\,3368 & NGC\,3941   \\
\hline
\hline
\multirow{6}{*}{Bulge}     & $\mu_e$          & 18.03 $\pm$ 0.26  & 18.21  $\pm$ 0.16  & 19.12   $\pm$ 0.26  & 16.68 $\pm$ 0.13  & 18.11 $\pm$ 0.16  & 16.66  $\pm$ 0.16  & 17.93  $\pm$ 0.15  & 18.30   $\pm$ 0.16  & 17.26  $\pm$ 0.13  \\   
                           & R$_e$          & 2.70  $\pm$ 0.43  & 3.38   $\pm$ 0.28  & 2.52    $\pm$ 0.40  & 2.59  $\pm$ 0.26  & 5.59  $\pm$ 0.47  & 2.90   $\pm$ 0.24  & 2.35   $\pm$ 0.24  & 13.75   $\pm$ 1.15  & 3.98   $\pm$ 0.40  \\ 
                           & n              & 1.63  $\pm$ 0.18  & 2.36   $\pm$ 0.18  & 1.62    $\pm$ 0.18  & 0.69  $\pm$ 0.05  & 1.66  $\pm$ 0.13  & 1.84   $\pm$ 0.14  & 1.61   $\pm$ 0.20  & 1.29    $\pm$ 0.10  & 2.62   $\pm$ 0.19  \\ 
                           & b/a            & 0.89  $\pm$ 0.04  & 0.84   $\pm$ 0.03  & 0.87    $\pm$ 0.04  & 0.96  $\pm$ 0.02  & 0.95  $\pm$ 0.03  & 0.72   $\pm$ 0.03  & 0.98   $\pm$ 0.04  & 0.70    $\pm$ 0.03  & 0.78   $\pm$ 0.02  \\ 
                           & PA             & 41.14 $\pm$ 5.05  & 165.64 $\pm$ 2.97  & 125.17  $\pm$ 5.05  & 0.00  $\pm$ 3.31  & 97.41 $\pm$ 2.97  & 97.97  $\pm$ 2.97  & 4.64   $\pm$ 5.20  & 156.86  $\pm$ 2.97  & 16.72  $\pm$ 3.31  \\ 
                           & B/T            & 0.164             & 0.209              & 0.085               & 0.129             & 0.237             & 0.267              & 0.135              & 0.197               & 0.226              \\ 
\hline                                                                                                                                                                                                                                                                                                                                                                                                              
\multirow{7}{*}{Disc}      & $\mu_o$          & 20.30 $\pm$ 0.07  & 20.13  $\pm$ 0.07  & 20.71   $\pm$ 0.07  & 19.77 $\pm$ 0.07  & 21.44 $\pm$ 0.07  & 19.73  $\pm$ 0.07  & 20.78  $\pm$ 0.07  & 20.22   $\pm$ 0.07  & 18.93  $\pm$ 0.07  \\ 
                           & h$_{\rm inner}$ & 30.02 $\pm$ 1.40  & 22.83  $\pm$ 0.38  & 30.91   $\pm$ 1.44  & 33.65 $\pm$ 0.81  & 67.91 $\pm$ 1.13  & 25.83  $\pm$ 0.43  & 37.46  $\pm$ 1.43  & 82.42   $\pm$ 1.37  & 32.08  $\pm$ 0.78  \\
                           & b/a            & 0.84  $\pm$ 0.01  & 0.95   $\pm$ 0.01  & 0.97    $\pm$ 0.01  & 0.89  $\pm$ 0.01  & 0.80  $\pm$ 0.01  & 0.65   $\pm$ 0.01  & 0.55   $\pm$ 0.01  & 0.67    $\pm$ 0.01  & 0.65   $\pm$ 0.01  \\ 
                           & PA             & 20.91 $\pm$ 0.37  & 16.23  $\pm$ 0.45  & 153.14  $\pm$ 0.37  & 12.22 $\pm$ 0.25  & 80.90 $\pm$ 0.45  & 124.19 $\pm$ 0.45  & 5.98   $\pm$ 0.58  & 158.78  $\pm$ 0.45  & 9.15   $\pm$ 0.25  \\ 
                           & h$_{\rm outer}$ & 15.39 $\pm$ 0.54  & N/A                & 14.20   $\pm$ 0.50  & 22.13 $\pm$ 0.77  & N/A               & N/A                & N/A                & N/A                 & 20.62  $\pm$ 0.72  \\
                           & R$_{\rm break}$ & 46.09 $\pm$ 1.56  & N/A                & 40.09   $\pm$ 1.36  & 88.81 $\pm$ 3.01  & N/A               & N/A                & N/A                & N/A                 & 38.39  $\pm$ 1.30  \\
                           & D/T            & 0.681             & 0.658              & 0.822               & 0.637             & 0.575             & 0.457              & 0.594              & 0.539               & 0.616              \\                       
\hline                                                                                                                                                                                                                                                                                                                                                                                                              
\multirow{7}{*}{Outer bar} & $\mu_o$          & 20.22 $\pm$ 0.30  & 20.02  $\pm$ 0.20  & 20.86   $\pm$ 0.30  & 19.21 $\pm$ 0.18  & 20.25 $\pm$ 0.20  & 19.28  $\pm$ 0.20  & 20.20  $\pm$ 0.32  & 19.56   $\pm$ 0.20  & 18.67  $\pm$ 0.18  \\ 
                           & a              & 38.09 $\pm$ 0.62  & 45.97  $\pm$ 0.50  & 48.35   $\pm$ 0.79  & 40.37 $\pm$ 0.40  & 72.16 $\pm$ 0.78  & 45.74  $\pm$ 0.50  & 49.62  $\pm$ 0.82  & 130.17  $\pm$ 1.41  & 35.13  $\pm$ 0.34  \\ 
                           & n              & 3                 & 4                  & 3                   & 3                 & 3                 & 3                  & 2                  & 3                   & 2                  \\ 
                           & b/a            & 0.44  $\pm$ 0.04  & 0.43   $\pm$ 0.03  & 0.26    $\pm$ 0.04  & 0.74  $\pm$ 0.02  & 0.63  $\pm$ 0.03  & 0.53   $\pm$ 0.03  & 0.49   $\pm$ 0.02  & 0.59    $\pm$ 0.03  & 0.44   $\pm$ 0.02  \\ 
                           & PA             & 119.74$\pm$ 0.28  & 150.81 $\pm$ 0.13  & 118.57  $\pm$ 0.28  & 84.79 $\pm$ 0.16  & 159.38$\pm$ 0.13  & 156.76 $\pm$ 0.13  & 173.62 $\pm$ 0.33  & 128.00  $\pm$ 0.13  & 171.50 $\pm$ 0.16  \\ 
                           & c              & 3                 & 2                  & 4                   & 2                 & 2                 & 2                  & 2                  & 2                   & 2                  \\ 
                           & Bar/T          & 0.101             & 0.120              & 0.089               & 0.159             & 0.169             & 0.195              & 0.225              & 0.245               & 0.131              \\ 
\hline                                                                                                                                                                                                                                                                                                                                                                                                              
\multirow{7}{*}{Inner bar} & $\mu_o$          & 19.40 $\pm$ 0.98  & 18.11  $\pm$ 1.06  & 18.01   $\pm$ 0.98  & 18.35 $\pm$ 1.09  & 17.92 $\pm$ 1.06  & 18.05  $\pm$ 1.06  & 19.55  $\pm$ 0.24  & 16.15   $\pm$ 1.06  & 18.58  $\pm$ 1.09  \\ 
                           & a              & 13.40 $\pm$ 2.42  & 4.16   $\pm$ 0.43  & 3.19    $\pm$ 0.58  & 18.52 $\pm$ 1.57  & 11.15 $\pm$ 1.14  & 17.16  $\pm$ 1.76  & 15.48  $\pm$ 2.62  & 8.88    $\pm$ 0.91  & 13.23  $\pm$ 1.12  \\ 
                           & n              & 3                 & 2                  & 2                   & 3                 & 3                 & 4                  & 3                  & 3                   & 3                  \\ 
                           & b/a            & 0.89  $\pm$ 0.06  & 0.63   $\pm$ 0.04  & 0.16    $\pm$ 0.06  & 0.75  $\pm$ 0.03  & 0.34  $\pm$ 0.04  & 0.61   $\pm$ 0.04  & 0.72   $\pm$ 0.06  & 0.41    $\pm$ 0.04  & 0.76   $\pm$ 0.03  \\ 
                           & PA             & 157.67$\pm$ 2.85  & 35.03  $\pm$ 1.30  & 159.41  $\pm$ 2.85  & 71.40 $\pm$ 2.09  & 61.78 $\pm$ 1.30  & 141.49 $\pm$ 1.30  & 0.80   $\pm$ 2.21  & 132.61  $\pm$ 1.30  & 170.13 $\pm$ 2.09  \\ 
                           & c              & 2                 & 4                  & 2                   & 2                 & 4                 & 2                  & 2                  & 4                   & 2                  \\ 
                           & Bar/T          & 0.054             & 0.013              & 0.004               & 0.075             & 0.019             & 0.081              & 0.046              & 0.018               & 0.027              \\   
\hline
\end{tabular}
}
\end{table}

\begin{table}
  \caption{ALL PARAMETERS FOR THE DOUBLE-BARRED FITS IN i'-BAND. Last set of 8/17 galaxies}
  \label{tab:paramsi3}
\resizebox{16.4cm}{!}{
  \begin{tabular}{llccccccccc}
  \hline
& & NGC\,3945 & NGC\,4314 & NGC\,4340 & NGC\,4503 & NGC\,4725 & NGC\,5850 & NGC\,7280 & NGC\,7716 \\
\hline
\hline
\multirow{6}{*}{Bulge}     & $\mu_e$          & 19.23   $\pm$ 0.16  & 19.23   $\pm$ 0.13 & 18.90  $\pm$ 0.16  & 18.62  $\pm$ 0.16  & 18.25  $\pm$ 0.16  & 19.77   $\pm$ 0.16  & 18.69  $\pm$ 0.26  & 17.54  $\pm$ 0.15  \\ 
                           & R$_e$          & 10.06   $\pm$ 0.84   & 11.25   $\pm$ 1.13 & 6.45   $\pm$ 0.51  & 5.39   $\pm$ 0.45  & 7.22   $\pm$ 0.61  & 8.79    $\pm$ 0.74  & 4.04   $\pm$ 0.65  & 1.97   $\pm$ 0.20  \\ 
                           & n              & 3.30    $\pm$ 0.26   & 1.00    $\pm$ 0.07 & 2.34   $\pm$ 0.17  & 2.70   $\pm$ 0.21  & 2.27   $\pm$ 0.18  & 2.46    $\pm$ 0.19  & 3.07   $\pm$ 0.33  & 2.19   $\pm$ 0.27  \\ 
                           & b/a            & 0.87    $\pm$ 0.03   & 0.81    $\pm$ 0.02 & 0.76   $\pm$ 0.03  & 0.76   $\pm$ 0.03  & 0.84   $\pm$ 0.03  & 0.97    $\pm$ 0.03  & 0.66   $\pm$ 0.04  & 0.57   $\pm$ 0.04  \\ 
                           & PA             & 147.98  $\pm$ 2.97   & 132.45  $\pm$ 3.31 & 98.75  $\pm$ 2.97  & 19.61  $\pm$ 2.97  & 28.52   $\pm$ 2.97  & 160.29  $\pm$ 2.97  & 68.06  $\pm$ 5.05  & 56.16  $\pm$ 5.20  \\ 
                           & B/T            & 0.329                & 0.210              & 0.241              & 0.207              & 0.101              & 0.195               & 0.269              & 0.154              \\ 
\hline                                                                                                                                                                                                                                                                                                                                                                                                                 
\multirow{7}{*}{Disc}      & $\mu_o$          & 21.12   $\pm$ 0.07  & 21.05   $\pm$ 0.07 & 20.33  $\pm$ 0.07  & 19.02  $\pm$ 0.07  & 19.59  $\pm$ 0.07  & 21.40   $\pm$ 0.07  & 20.19  $\pm$ 0.07  & 19.73  $\pm$ 0.07  \\ 
                           & h$_{\rm inner}$ & 57.57   $\pm$ 0.95     & 63.47   $\pm$ 1.53 & 35.24  $\pm$ 0.58  &26.62   $\pm$ 0.44  & 85.69   $\pm$ 1.42  & 66.96   $\pm$ 1.11  & 27.34  $\pm$ 1.27  & 15.95  $\pm$ 0.61  \\
                           & b/a            & 0.68    $\pm$ 0.01   & 0.92    $\pm$ 0.01 & 0.69   $\pm$ 0.01  & 0.46   $\pm$ 0.01  & 0.47   $\pm$ 0.01  & 0.75    $\pm$ 0.01  & 0.65   $\pm$ 0.01  & 0.83   $\pm$ 0.01  \\ 
                           & PA             & 163.30  $\pm$ 0.45    & 136.98  $\pm$ 0.25 & 91.44  $\pm$ 0.45  & 9.79   $\pm$ 0.45  & 45.33  $\pm$ 0.45  & 139.03  $\pm$ 0.45  & 74.00  $\pm$ 0.37  & 40.91  $\pm$ 0.58  \\ 
                           & h$_{\rm outer}$ & N/A                    & 28.52   $\pm$ 1.00 & N/A                & N/A                & N/A                & N/A                 & 11.92  $\pm$ 0.42  & N/A                \\
                           & R$_{\rm break}$ & N/A                    & 102.74  $\pm$ 3.48 & N/A                & N/A                & N/A                & N/A                 & 44.19  $\pm$ 1.50  & N/A                \\
                           & D/T            & 0.450                 & 0.498              & 0.621              & 0.711              & 0.854              & 0.683               & 0.636              & 0.719              \\ 
\hline                                                                                                                                                                                                                                                                                                                                                                                                                 
\multirow{7}{*}{Outer bar} & $\mu_o$          & 20.11   $\pm$ 0.20  & 19.87   $\pm$ 0.18 & 21.31  $\pm$ 0.20  & 19.98  $\pm$ 0.20  & 19.87  $\pm$ 0.20  & 21.08   $\pm$ 0.20  &  20.23 $\pm$ 0.30  & 20.89  $\pm$ 0.32  \\ 
                           & a              & 58.81   $\pm$ 0.64  & 126.15  $\pm$ 1.24 & 62.71  $\pm$ 0.83  & 32.83  $\pm$ 0.36  & 56.18 $\pm$ 0.61  & 134.14  $\pm$ 1.45  & 20.69  $\pm$ 0.34  & 28.85  $\pm$ 0.48  \\ 
                           & n              & 2                   & 3                  & 1                  & 2                  & 3                  & 4                   & 2                  & 2                  \\ 
                           & b/a            & 0.61    $\pm$ 0.03  & 0.25    $\pm$ 0.02 & 0.44   $\pm$ 0.03  & 0.45   $\pm$ 0.03  & 0.60   $\pm$ 0.03  & 0.25    $\pm$ 0.03  & 0.49   $\pm$ 0.04  & 0.51   $\pm$ 0.02  \\ 
                           & PA             & 72.26   $\pm$ 0.13  & 146.89  $\pm$ 0.16 & 30.93  $\pm$ 0.13  & 176.77 $\pm$ 0.13  & 34.57  $\pm$ 0.13  & 115.75  $\pm$ 0.13  & 55.17  $\pm$ 0.28  & 16.61  $\pm$ 0.33  \\ 
                           & c              & 2                   & 2                  & 4                  & 2                  & 3                  & 2                   & 2                  & 3                  \\ 
                           & Bar/T          & 0.153               & 0.265              & 0.120              & 0.063              & 0.040              & 0.112               & 0.056              & 0.072              \\ 
\hline                                                                                                                                                                                                                                                                                                                                                                                                                 
\multirow{7}{*}{Inner bar} & $\mu_o$          & 18.05   $\pm$ 1.06  & 16.35   $\pm$ 1.09 & 18.03  $\pm$ 1.06  & 18.88  $\pm$ 1.06  & 17.98  $\pm$ 1.06  & 18.75   $\pm$ 1.06  & 14.61  $\pm$ 0.98  & 18.69  $\pm$ 0.24  \\ 
                           & a              & 19.78   $\pm$ 2.03   & 4.80    $\pm$ 0.41 & 8.88   $\pm$ 0.91  & 10.54  $\pm$ 1.08  & 12.30  $\pm$ 1.26  & 11.15   $\pm$ 1.14  & 2.23   $\pm$ 0.40  &  8.12  $\pm$ 1.37  \\ 
                           & n              & 3                    & 4                  & 4                  & 4                  & 3                  & 2                   & 2                  & 3                  \\ 
                           & b/a            & 0.46    $\pm$ 0.04   & 0.82    $\pm$ 0.03 & 0.36   $\pm$ 0.04  & 0.75   $\pm$ 0.04  & 0.28   $\pm$ 0.04  & 0.24    $\pm$ 0.04  & 0.17   $\pm$ 0.06  & 0.84   $\pm$ 0.06  \\ 
                           & PA             & 159.84  $\pm$ 1.30   & 152.92  $\pm$ 2.09 & 19.82  $\pm$ 1.30  & 150.13 $\pm$ 1.30  & 137.08  $\pm$ 1.30  & 47.49   $\pm$ 1.30  & 112.04 $\pm$ 2.85  & 53.15  $\pm$ 2.21  \\ 
                           & c              & 3                    & 2                  & 4                  & 2                  & 5                  & 4                   & 1                 & 2                  \\ 
                           & Bar/T          & 0.067                & 0.026              & 0.019              & 0.019              & 0.005              & 0.010               & 0.040              & 0.055              \\ 
\hline
\end{tabular}
}
\end{table}

\begin{table}
 \centering
  \caption{ALL PARAMETERS FOR THE SINGLE-BARRED FITS IN r'-BAND. First set of 9/17 galaxies.}
  \label{tab:paramsone1}
\resizebox{18cm}{!}{
  \begin{tabular}{llccccccccc}
  \hline
& & \multicolumn{1}{c}{NGC\,357} & \multicolumn{1}{c}{NGC\,718} & \multicolumn{1}{c}{NGC\,2642} & NGC\,2681 & NGC\,2859 & NGC\,2950 & NGC\,2962 & NGC\,3368 & NGC\,3941  \\
\hline
\hline
\multirow{6}{*}{Bulge}     & $\mu_e$          & 18.72  $\pm$ 0.10  & 17.98  $\pm$ 0.17  & 19.45  $\pm$ 0.19  &  16.96 $\pm$ 0.07  &  18.10  $\pm$ 0.09  & 17.26  $\pm$ 0.09 & 18.70  $\pm$ 0.17  & 18.95   $\pm$ 0.09  & 17.66  $\pm$ 0.07 \\ 
                           & R$_e$          & 3.46   $\pm$ 0.23  & 2.62   $\pm$ 0.33  & 2.53   $\pm$ 0.30  &  2.99  $\pm$ 0.13  &  5.01   $\pm$ 0.26  & 3.68   $\pm$ 0.19 & 3.32   $\pm$ 0.42  & 15.18   $\pm$ 0.79  & 4.55   $\pm$ 0.20 \\ 
                           & n              & 1.63   $\pm$ 0.09  & 1.93   $\pm$ 0.14  & 1.71   $\pm$ 0.19  &  1.65  $\pm$ 0.06  &  1.45   $\pm$ 0.06  & 2.03   $\pm$ 0.09 & 1.80   $\pm$ 0.13  & 1.94    $\pm$ 0.09  & 2.27   $\pm$ 0.08 \\ 
                           & b/a            & 0.93   $\pm$ 0.02  & 0.86   $\pm$ 0.03  & 0.82   $\pm$ 0.04  &  0.92  $\pm$ 0.01  &  0.87   $\pm$ 0.02  & 0.74   $\pm$ 0.02 & 0.90   $\pm$ 0.03  & 0.68    $\pm$ 0.02  & 0.79   $\pm$ 0.01 \\ 
                           & PA             & 27.46  $\pm$ 3.06  & 172.88 $\pm$ 2.80  & 134.82 $\pm$ 5.23  &  52.81 $\pm$ 1.87  &  72.40  $\pm$ 3.05  & 110.61 $\pm$ 3.05 & 179.38 $\pm$ 2.80  & 151.07  $\pm$ 3.05  & 12.81  $\pm$ 1.87 \\ 
                           & B/T            & 0.245              & 0.211              & 0.097              &  0.255             &  0.268              & 0.389             & 0.207              & 0.251               & 0.282             \\ 
\hline                                                                                                                                                                                                                                                                                                                                                                                                       
\multirow{7}{*}{Disc}      & $\mu_o$          & 20.96  $\pm$ 0.05  & 20.49  $\pm$ 0.05  & 21.11  $\pm$ 0.11  &  20.14 $\pm$ 0.11  &  21.73  $\pm$ 0.11  & 20.08  $\pm$ 0.11 & 21.10  $\pm$ 0.05  & 20.77   $\pm$ 0.05  & 19.42  $\pm$ 0.05 \\ 
                           & h$_{\rm inner}$ & 33.25  $\pm$ 1.15  & 22.06  $\pm$ 0.86  & 27.61  $\pm$ 2.04  &  33.26 $\pm$ 0.68  &  56.99  $\pm$ 0.84  & 25.33  $\pm$ 0.37 & 33.42  $\pm$ 1.30  & 82.14   $\pm$ 1.21  & 34.14  $\pm$ 0.69 \\
                           & b/a            & 0.82   $\pm$ 0.01  & 0.95   $\pm$ 0.02  & 0.97   $\pm$ 0.02  &  0.89  $\pm$ 0.01  &  0.85   $\pm$ 0.01  & 0.64   $\pm$ 0.01 & 0.57   $\pm$ 0.02  & 0.65    $\pm$ 0.01  & 0.64   $\pm$ 0.01 \\ 
                           & PA             & 18.58  $\pm$ 3.61  & 16.97  $\pm$ 3.92  & 151.36 $\pm$ 7.19  &  12.33 $\pm$ 1.31  &  85.66  $\pm$ 1.98  & 125.91 $\pm$ 1.98 & 2.38   $\pm$ 3.92  & 161.32  $\pm$ 1.98  & 9.21   $\pm$ 1.31 \\ 
                           & h$_{\rm outer}$ & 14.14  $\pm$ 0.71  & N/A                & 12.01  $\pm$ 0.60  &  18.66 $\pm$ 0.93  &  N/A                & N/A                & N/A                & N/A                 & 20.15  $\pm$ 1.01 \\
                           & R$_{\rm break}$ & 47.01  $\pm$ 1.35  & N/A                & 45.54  $\pm$ 1.31  &  88.63 $\pm$ 2.54  &  N/A                & N/A               & N/A                & N/A                 & 39.49  $\pm$ 1.13 \\
                           & D/T            & 0.652              & 0.651              & 0.806              &  0.595             &  0.537              & 0.453             & 0.588              & 0.516               & 0.595             \\ 
\hline                                                                                                                                                                                                                                                                                                                                                                                                       
\multirow{7}{*}{Outer bar} & $\mu_o$          & 20.70  $\pm$ 0.31  & 20.29  $\pm$ 0.25  & 21.36  $\pm$ 0.57  &  19.59 $\pm$ 0.24  &  20.53  $\pm$ 0.16  & 19.75  $\pm$ 0.16 & 20.79  $\pm$ 0.25  & 20.23   $\pm$ 0.16  & 19.11  $\pm$ 0.24 \\ 
                           & a              & 37.34  $\pm$ 0.74  & 45.33  $\pm$ 1.12  & 52.93  $\pm$ 2.51  &  40.96 $\pm$ 0.46  &  69.60  $\pm$ 0.80  & 46.36  $\pm$ 0.53 & 50.83  $\pm$ 1.26  & 136.20  $\pm$ 1.56  & 35.93  $\pm$ 0.41 \\ 
                           & n              & 3                  & 4                  & 3                  &  3                 &  3                  & 3                 & 2                  & 3                   & 2                 \\ 
                           & b/a            & 0.44   $\pm$ 0.03  & 0.44   $\pm$ 0.06  & 0.24   $\pm$ 0.07  &  0.69  $\pm$ 0.02  &  0.62   $\pm$ 0.03  & 0.44   $\pm$ 0.03 & 0.45   $\pm$ 0.06  & 0.59    $\pm$ 0.03  & 0.41   $\pm$ 0.02 \\ 
                           & PA             & 119.74 $\pm$ 3.02  & 150.94 $\pm$ 2.00  & 116.85 $\pm$ 4.79  &  84.81 $\pm$ 1.64  &  159.22 $\pm$ 1.89  & 158.60 $\pm$ 1.89 & 174.36 $\pm$ 2.00  & 127.74  $\pm$ 1.89  & 171.24 $\pm$ 1.64 \\ 
                           & c              & 3                  & 2                  & 4                  &  2                 &  2                  & 2                 & 2                  & 2                   & 2                 \\ 
                           & Bar/T          & 0.103              & 0.138              & 0.097              &  0.149             &  0.195              & 0.159             & 0.205              & 0.233               & 0.123             \\ 
\hline
\end{tabular}
}
\end{table}

\begin{table}
  \caption{ALL PARAMETERS FOR THE SINGLE-BARRED FITS IN r'-BAND. Last set of 8/17 galaxies}
  \label{tab:paramsone3}
\resizebox{16.4cm}{!}{
  \begin{tabular}{llccccccccccc}
  \hline
& & NGC\,3945 & NGC\,4314 & NGC\,4340 & NGC\,4503 & NGC\,4725 & NGC\,5850 & NGC\,7280 & NGC\,7716 \\
\hline
\hline
\multirow{6}{*}{Bulge}     & $\mu_e$          & 18.08  $\pm$ 0.09   & 19.72   $\pm$ 0.07  & 18.81  $\pm$ 0.09  & 18.64  $\pm$ 0.09  & 18.12  $\pm$ 0.09  & 19.61  $\pm$ 0.09  & 18.64  $\pm$ 0.10  & 18.07  $\pm$ 0.17  \\ 
                           & R$_e$          & 6.33   $\pm$ 0.33   & 10.95   $\pm$ 0.48  & 4.97   $\pm$ 0.26  & 4.64   $\pm$ 0.24  & 5.43  $\pm$ 0.28  & 6.82   $\pm$ 0.35  & 3.19   $\pm$ 0.21  & 2.45   $\pm$ 0.31  \\ 
                           & n              & 1.28   $\pm$ 0.06   & 2.07    $\pm$ 0.07  & 2.04   $\pm$ 0.09  & 2.23   $\pm$ 0.10  & 1.54   $\pm$ 0.07  & 1.99   $\pm$ 0.09  & 3.49   $\pm$ 0.19  & 1.52   $\pm$ 0.11  \\ 
                           & b/a            & 0.69   $\pm$ 0.02   & 0.81    $\pm$ 0.01  & 0.86   $\pm$ 0.02  & 0.79   $\pm$ 0.02  & 0.92   $\pm$ 0.02  & 0.89   $\pm$ 0.02  & 0.67   $\pm$ 0.02  & 0.67   $\pm$ 0.03  \\ 
                           & PA             & 156.67 $\pm$ 3.05   & 134.37  $\pm$ 1.87  & 91.60  $\pm$ 3.05  & 12.92  $\pm$ 3.05  & 5.76   $\pm$ 3.05  & 41.03  $\pm$ 3.05  & 72.10  $\pm$ 3.06  & 55.53  $\pm$ 2.80  \\ 
                           & B/T            & 0.317               & 0.261               & 0.259              & 0.219              & 0.088              & 0.179              & 0.288              & 0.201              \\ 
\hline                                                                                                                                                                                                                                                                                                                                                                                      
\multirow{7}{*}{Disc}      & $\mu_o$          & 21.12  $\pm$ 0.05   & 21.49   $\pm$ 0.05  & 20.68  $\pm$ 0.05  & 19.43  $\pm$ 0.05  & 20.05  $\pm$ 0.05  & 21.58  $\pm$ 0.05  & 20.59  $\pm$ 0.05  & 20.13  $\pm$ 0.05  \\ 
                           & h$_{\rm inner}$ &44.60   $\pm$ 0.66   & 62.51   $\pm$ 1.27   & 33.76   $\pm$ 0.50  &26.20   $\pm$ 0.39  & 85.06   $\pm$ 1.25  & 59.17  $\pm$ 0.87  & 26.95  $\pm$ 0.94  & 16.44  $\pm$ 0.64  \\
                           & b/a            & 0.72   $\pm$ 0.01   & 0.92    $\pm$ 0.01  & 0.66   $\pm$ 0.01  & 0.46   $\pm$ 0.01  & 0.50   $\pm$ 0.01  & 0.74   $\pm$ 0.01  & 0.65   $\pm$ 0.01  & 0.82   $\pm$ 0.02  \\ 
                           & PA             & 156.77 $\pm$ 1.98   & 137.39  $\pm$ 1.31  & 92.96  $\pm$ 1.98  & 9.99   $\pm$ 1.98  & 43.07  $\pm$ 1.98  & 136.37 $\pm$ 1.98  & 73.83  $\pm$ 3.60  & 41.36  $\pm$ 3.92  \\ 
                           & h$_{\rm outer}$ & N/A                 & 25.72   $\pm$ 1.29   & N/A                & N/A                & N/A                & N/A                & 11.19  $\pm$ 0.56  & N/A                \\
                           & R$_{\rm break}$ & N/A                 & 105.84  $\pm$ 3.03   & N/A                & N/A                & N/A                & N/A                & 44.40  $\pm$ 1.27  & N/A                \\
                           & D/T            & 0.468               & 0.480               & 0.628              & 0.722              & 0.864              & 0.706              & 0.639              & 0.732              \\ 
\hline                                                                                                                                                                                                                                                                                                                                                                                      
\multirow{7}{*}{Outer bar} & $\mu_o$          & 20.25  $\pm$ 0.16   & 20.25   $\pm$ 0.24  & 21.75  $\pm$ 0.16  & 20.51  $\pm$ 0.16  & 20.11  $\pm$ 0.16  &  21.41 $\pm$ 0.16  & 20.40  $\pm$ 0.31  & 21.43  $\pm$ 0.25  \\ 
                           & a              & 57.05  $\pm$ 0.65   & 123.78  $\pm$ 1.40  & 63.33  $\pm$ 0.73  & 33.19  $\pm$ 0.38  & 56.17 $\pm$ 0.64  & 123.71 $\pm$ 1.42  & 20.90  $\pm$ 0.42  & 32.07  $\pm$ 0.79  \\ 
                           & n              & 2                   & 3                   & 1                  & 2                  & 3                  & 4                  & 2                  & 2                  \\ 
                           & b/a            & 0.63   $\pm$ 0.03   & 0.24    $\pm$ 0.02  & 0.45   $\pm$ 0.03  & 0.45   $\pm$ 0.03  & 0.60   $\pm$ 0.03  & 0.26   $\pm$ 0.03  & 0.48   $\pm$ 0.03  & 0.46   $\pm$ 0.06  \\   
                           & PA             & 73.25  $\pm$ 1.89   & 146.78  $\pm$ 1.64  & 31.14  $\pm$ 1.89  & 177.80 $\pm$ 1.89  & 36.86  $\pm$ 1.89  & 116.54 $\pm$ 1.89  & 55.13  $\pm$ 3.02  & 16.95  $\pm$ 2.00  \\ 
                           & c              & 2                   & 2                   & 4                  & 2                  & 3                  & 2                  & 2                  & 3                  \\ 
                           & Bar/T          & 0.215               & 0.258               & 0.112              & 0.060              & 0.048              & 0.114              & 0.073              & 0.067              \\ 
\hline
\end{tabular}
}
\end{table}
\end{onecolumn}

\label{lastpage}

\end{document}